\documentclass[pdflatex,sn-mathphys-num]{sn-jnl}%
\usepackage{graphicx}%
\usepackage{multirow}%
\usepackage{amsmath,amssymb,amsfonts}%
\usepackage{amsthm}%
\usepackage{mathrsfs}%
\usepackage[title]{appendix}%
\usepackage{xcolor}%
\usepackage{textcomp}%
\usepackage{manyfoot}%
\usepackage{booktabs}%
\usepackage{algorithm}%
\usepackage{algorithmicx}%
\usepackage{algpseudocode}%
\usepackage{listings}%
\usepackage{graphicx}
\usepackage{subcaption}
\usepackage[capitalise,noabbrev]{cleveref}
\usepackage{float}
\providecommand{\doiurl}[1]{\href{https://doi.org/#1}{\nolinkurl{#1}}}

\begin{document}


\title[Article Title]{A well-motivated model of pedestrian dynamics}



\author*[1]{\fnm{Ezel} \sur{\"Usten}}\email{e.uesten@fz-juelich.de}

\author[1,3]{\fnm{Anna} \sur{Sieben}}

\author[1]{\fnm{Mohcine} \sur{Chraibi}}

\author[1,2]{\fnm{Armin} \sur{Seyfried}}

\affil*[1]{\orgdiv{Institute for Advanced Simulation (IAS)}, \orgname{Forschungszentrum J\"ulich}, \city{J\"ulich}, \country{Germany}}

\affil[2]{\orgdiv{Faculty of Architecture and Civil Engineering}, \orgname{University of Wuppertal}, \city{Wuppertal}, \country{Germany}}

\affil[3]{\orgdiv{Faculty of Humanities and Social Sciences}, \orgname{University of Wuppertal}, \city{Wuppertal}, \country{Germany}}


 \abstract{In pedestrian dynamics, the internal drive that propels individuals toward their goals is typically captured by a
  single, fixed parameter, the desired walking speed. 
  This simplification overlooks that motivation fluctuates in response to changing spatial and social conditions within a crowd. 
  This paper proposes a dynamic motivation model grounded in expectancy-value theory from psychology, in which each agent's motivation evolves over time depending on proximity to the goal, relative position among other pedestrians, and individual goal importance. 
  The resulting motivation modulates multiple movement parameters simultaneously, including walking speed, gap-closing behavior, and interpersonal spacing. 
  The model is evaluated in simulated pre-bottleneck waiting scenarios using paired statistical comparisons across
  multiple random seeds and population sizes, and compared with trajectory data from the CROMA concert-entry
  bottleneck experiments under low- and high-motivation framings.
  Simulations show that the dynamic model produces structured heterogeneity in the crowd: agents self-organize into differentiated positions near the bottleneck, with those closer to the front occupying less space, a pattern absent in the static baseline but clearly present in the experimental data.
  These findings suggest that motivation in crowds should be understood not as a uniform increase in urgency, but as a mechanism that reorganizes competitive positioning along spatial and social axes. 
  Future work should extend the framework to open-door throughput scenarios, larger populations, and richer social interactions such as group cohesion and cooperative strategies.}


\maketitle

\section{Introduction}\label{introduction}

In pedestrian dynamics, the propulsive drive of moving agents plays a fundamental role. The intensity with which agents accelerate, how fast they move, how closely they approach to one another, the pressure they exert on those in front of them in high density conditions, and their tendency to overtake others can all be interpreted as behavioral expressions of how strongly they are oriented toward their goal. In this article, we use the psychological concept of motivation to describe this propulsive drive. Originating from the Latin word \textit{movere} (``to move''), motivation is commonly defined as ``an internal process that energizes and directs behavior'' \cite[p.~9]{reeve2018}. In line with Lewin's field theory, we adopt a broader perspective that emphasizes the interaction between the individual and the environment: Motivation can be understood as the dynamic interplay of forces within the person–environment field that determines the direction, intensity, and persistence of behavior \cite{lewin_field_1942}.

In computational models of pedestrian dynamics, this goal-directed drive is typically operationalized through movement parameters. All agent-based models in pedestrian dynamics can, and some explicitly do, represent motivational differences in a simple manner. 
The parameter most commonly used to reflect motivational level is the free speed $v_0$, also referred to as intended, desired, or preferred speed. The underlying assumption here is that high motivation goes hand in hand with a high preferred speed. This parameter can be specified in two principal ways: First, all agents can be uniformly assigned high or low intended speeds, allowing the model to distinguish between situations in which everyone wants to move quickly (i.e., rushing to catch a departing train) and situations in which speed is of little relevance (i.e. the train has not yet arrived). Second, heterogeneity between agents can be introduced by implementing a distribution of intended speeds across agents, thereby, the simulation takes individual differences in motivational level into account. In both cases the motivation remains static, either fixed for the entire simulation or fixed at the individual level. We therefore refer to this as a static approach to modeling motivation.

The dynamic motivation model proposed in this paper extends this approach in two ways. First, it assumes that motivation is an underlying psychological construct that influences not only intended speed but most (if not all) parameters of a movement model. These include interpersonal distance, the frequency and intensity of directional changes in movement, reaction times, and acceleration or speed changes. Second, the motivation is treated as dynamically evolving, depending on an individual's position relative to the spatial surroundings (i.e. the entrance) and to the other pedestrians. Consider the following two examples: A person on the way to a concert entrance may initially be highly motivated to enter early. If she is repeatedly overtaken by others and perceives that reaching the front is unlikely, eventually her motivation will decrease because it does not seem worth trying to be one of the first anymore. 
Conversely, a person waiting at the end of a very long queue at the supermarket checkout may show little motivation to move fast or close gaps to the person in front. If a second checkout opens, the same individual may suddenly become very motivated to be one of the first in the queue. We formalize these dynamic changes in motivation using the psychological expectancy-value model. The motivation value determined using this model is dynamically updated over time and then incorporated into the key parameters of our movement model.

With this model, we pursue three objectives, which are reflected in the following three parts of the introduction. First, the model is based on empirical findings from research on crowd dynamics. Accordingly, the first part reviews how motivation has been integrated into previous studies and what behavioral effects have been demonstrated. Second, the model is intended to be psychologically sound. The second part therefore introduces the psychological expectancy value theory and adapts it to the context of pedestrian movement. Third, the model is based on state-of-the-art pedestrian modeling approaches. The third part places the proposed framework in the context of these movement models.

\subsection{Part One: Motivation in empirical studies} 

This section discusses the effects of motivation across a variety of settings in pedestrian dynamics. It mainly focuses on controlled pedestrian experiments rather than field studies. It should be noted, however, that although the effects of motivation have been frequently described in the literature, the term ``motivation'' itself has not always been used explicitly. We will address this further below. 

Most empirical studies in the field that explore the role of motivation do so in the context of bottlenecks as the primary spatial structure. 
Such bottlenecks are not accidental; they are intentionally integrated into modern public infrastructure, commonly implemented at concerts, event venues, building exits, and transportation hubs to enable access control or for reasons of structural efficiency. 
Beyond this everyday role, bottlenecks also appear in a safety-critical form as emergency exits, where the same geometry constraints evacuation rather than  throughput. 
These environments, now integral to everyday life, create settings in which personal motivation interacts with physical limitations and social context, resulting in observable movement patterns. Only a limited number of studies have investigated motivational influences in other contexts, such as corridors with different densities \cite{ye_pedestrian_2021, ziemer_mikroskopische_2020} or staircases \cite{graat_complex_1999}. 

The focus on bottlenecks is closely linked to the well-known effect of clogging and the ``faster-is-slower effect''. However, despite its widespread use, the term requires cautious and critical examination, as the phenomenon occurs only under very specific conditions. 
In a relevant article, Muir et al. \cite{muir_effects_1996} demonstrated in an aircraft evacuation study that a high motivation lead to shorter evacuation times (higher flow) when the bottleneck is wide. In contrast, narrow bottlenecks tended to trigger temporary clogs, resulting in longer evacuation times (lower flow). Further studies examining the effect of motivation on clogging probability, flow rate reduction, or, equivalently, the extension of the clearance time, can be found in \cite{muller_fluchtwegen_1981,pastor_experimental_2015,garcimartin_flow_2016,li_bottleneck_2020,li_motivation_2024,Helbing2005b}. In summary, these findings show that a reduction of the flow due to clogging only occurs when bottleneck width is below approximately 80 cm and only under conditions of very high motivation. This effect is commonly referred to as the ``faster-is-slower effect'' effect. Its practical relevance should therefore be carefully considered; a point we return to later in this article.

In addition to the flow rate, the effects of high motivation are also evident in pre-bottleneck density: High motivation groups tend to show significantly higher density in front of the bottleneck \cite{sieben_collective_2017, adrian_crowds_2020, li_bottleneck_2021}.  

A closer look at the experimental designs of these studies shows that they use various methods and instructions to increase or vary the motivation level of the participants. Some studies manipulated motivation through explicit instructions about movement speed (e.g., ``normal'', ``fast'', or ``as fast as possible'', \cite{peschl_passage_1971, muller_fluchtwegen_1981}), or through directives regarding physical contact (e.g. ``no contact'', ``only accidental contact'', ``contact allowed'', or ``pushing allowed'' \cite{garcimartin_flow_2016, pastor_experimental_2015}). In the bottleneck experiments described in Boomers at al. \cite{boomers_crowd_2023}, both speed and contact instructions are systematically varied across three motivational levels. Other studies manipulate motivation more implicitly by creating context specific scenarios, such as an emergency evacuation. Some experiments are conducted without any additional incentive \cite{li_bottleneck_2020, li_bottleneck_2021}, while others introduce a reward structure alongside the scenario. In these experiments, participants receive an additional sum $X$ if they are among the first $m$ out of $N$ individuals to pass through the bottleneck \cite{mintz_nonadaptive_1951, muir_effects_1996, li_motivation_2024}. Other studies create a contextual motivation through imaginative framing, such as attending a concert \cite{sieben_collective_2017, adrian_crowds_2020, sieben_repertoires_2025}. This approach relies entirely on the imagination of the participants, using instructions such as: 
\begin{quote}
\textit{``Imagine you are attending a concert by your favorite artist and are waiting in line to get in.''}
\end{quote}

It is worth noting that related constructs (such as competitiveness and cooperativeness) frequently appear in the literature and are sometimes equated with motivation \cite{mintz_nonadaptive_1951, sime_escape_1983}. Some authors assume that high motivation implies competitive behavior. A common example can be found in evacuation scenarios, where it is often presumed that individuals, being highly motivated to exit quickly, will therefore act selfishly and competitively. However, this assumption has been proven inaccurate by several studies that reconstruct behavior during emergencies, showing instead that cooperation often occurs \cite{drury_collective_2009, levine_identity_2005}. We argue that  cooperation or competition are not universal indicators of motivational strength but instead depend on social norms of the setting and strategic considerations. In contexts such as sports (i.e., race), for example, high motivation is inherently linked to competition (i.e., to be the first, winning), whereas in other settings, strategic cooperation may come with high motivation. The role of strategic decisions leading to either cooperation or competition has also been extensively studied in the game theory research \cite{axelrod_evolution_1981}. 

\begin{figure}
    \centering
    \includegraphics[width=1.0\linewidth]{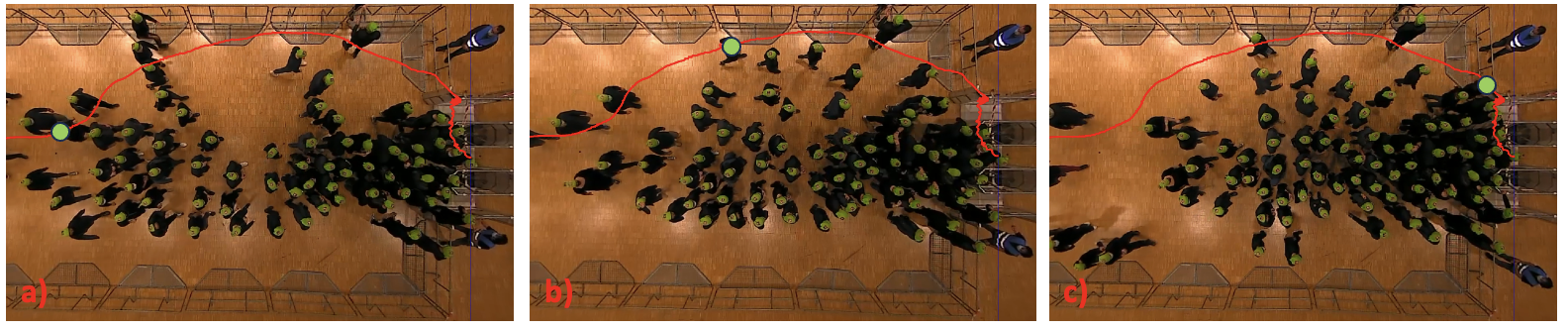}
    \caption{Three snapshots from an crowd experiment in which participants were instructed to imagine being on the way to a concert of their favorite artist  \cite{boomers_crowd_2023, sieben_repertoires_2025}. One trajectory is highlighted to follow person x. In a) person x (green circle) enters the measuring area and finds the space in front of the entrance crowded. In b) the person x takes a long detour and runs around the crowd. in c) the person x takes a waiting position very close to the entrance gate. This exemplifies the heterogeneity of individual behavior even in controlled experiments.}
    \label{fig:heterogeneity}
\end{figure}

In addition to this conceptual dichotomy, experimental research often relies on a further simplifying assumption. Most studies presented so far implicitly assume that all participants in the ``high motivation'' condition would behave uniformly, in a way that collectively represented a high-motivation group. However, in real-world scenarios, crowd motivation is not monolithic. Individuals may vary in their motivation: some may be highly driven to reach to the front (e.g., to secure a spot or leave quickly), while others may be indifferent, preferring to arrive later or simply follow the crowd. Interestingly, even in the controlled settings of experiments, participants do not behave uniformly. In our own experiments, we frequently observed that some participants appeared to treat the experiment as a game, displaying high motivation and assertive movement; others simply followed instructions without extra effort; and some disengaged entirely, remaining at the back and avoiding the denser parts of the crowd (see Figure ~\ref{fig:heterogeneity}, also see Figures 6 and 7 in \cite{adrian_crowds_2020}). This behavioral variability reflects the underlying heterogeneity of crowd composition, even in intentionally uniform and homogeneous experimental setups. 

Finally, some studies also show that motivation changes dynamically over time and space. In real-world entrance scenarios involving large numbers of people (such as a concert or festival) an individual's motivation can fluctuate due to various situational factors: how far they are from the entrance, how many people are ahead of them, or whether they believe they will reach the exit  relatively on time. Motivation may increase as the goal becomes more attainable (as suggested by the goal proximity hypothesis in consumer behavior \cite{jhang_impatience_2015}), or decrease when the situation seems less likely (e.g., when the density ahead is too high). Accordingly, in many bottleneck experiments a heterogenous structure is observable, with high densities in the front and densities decreasing toward the rear (\cite{sieben_collective_2017, adrian_crowds_2020}).

\begin{figure}
    \centering
    \includegraphics[width=0.5\linewidth]{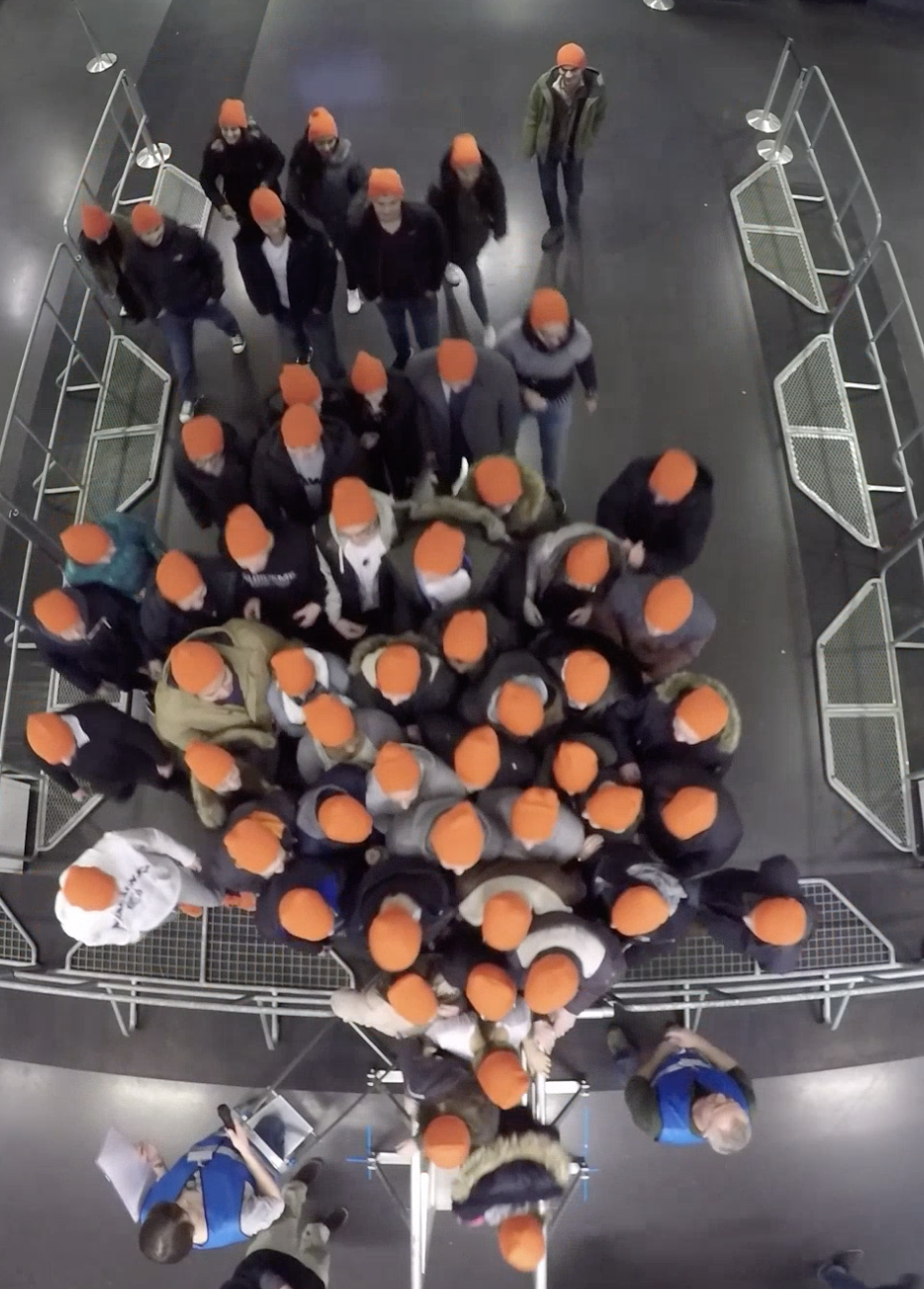}
    \caption{Still image of a crowd experiment in which participants were again instructed to imagine being on the way to a concert of their favorite artist \cite{adrian_crowds_2020}. The image shows a typical distribution of behavior and densities: Close to the gate people are pushing and density is high. Density gradually decreases toward the end of the crowds. The participants in the margins of the crowd leave gaps in front of them and some display ``unmotivated'' behavior. }
    \label{fig:experiment_still}
\end{figure}

This dynamism can be observed and quantified. Our research has demonstrated that individual behavior variations can be meaningfully captured by coding movement intensity. We developed a four-level ordinal scale and rated individuals accordingly for each time unit available: \textit{falling behind, just walking, mild pushing, and strong pushing} (ranging from least to most intense) \cite{usten_pushing_2022}. 
The results indicated that pedestrian behavior varied considerably over time and space; individuals rarely remained fixed in a single category, and the distribution of behaviors was broadly heterogeneous, even though participants were uniformly instructed to adopt either high or low motivation (see Figure ~\ref{fig:experiment_still}). 
Moreover, within this behavioral variability, distinct patterns emerged. 
Forward motion intensity appears to increase as individuals approach the bottleneck, reinforcing the idea that motivation is dynamically shaped by both spatial and social context \cite{usten_dynamic_2023}. 

Taken together, this body of research and conceptual groundwork points to a crucial insight: motivation in crowd scenarios is characterized, first, by the overarching (and static) context, for example, whether the situation involves an evacuation, an exit from a festival, or entry during Black Friday shopping. 
At the same time, motivation in crowd is also shaped by both heterogeneity between individuals and temporal and spacial dynamism within individuals. 
Any effort to model pedestrian motivation must therefore move beyond static assumptions and incorporate a psychological framework capable of capturing moment-to-moment motivational change across pedestrians.

The empirical reference for the present work is the bottleneck experiment by Adrian et al.~\cite{adrian_crowds_2020} and Sieben et al.~\cite{sieben_repertoires_2025}, in which participants were tasked to imagine waiting in front of a concert venue and their motivation is varied through the framing of the seating instructions. We use these datasets throughout the paper as the empirical counterpart to our simulations; the experimental setup is summarized in Section~\ref{sec:methods-experiment}.

\subsection{Part Two: A psychological framework of motivation}
\label{sec:part2}
Motivation has long been a central topic in psychology, with influential theories developed across domains such as basic needs, academic performance, occupational achievement, and long-term goal pursuit (e.g., \cite{lewin_field_1942,atkinson_introduction_1964,vroom_work_1964}). 
These frameworks have been foundational in advancing our understanding of behavior, but they predominantly address long-term or domain-specific patterns (career development, academic achievement, or interpersonal relationships). 
In contrast, pedestrian movement unfolds over seconds, not seasons. 
It involves highly situated, moment-to-moment decisions within dynamic physical and social environments, often under time pressure and spatial constraint. Despite over 80 years of motivational research, this type of short-timescale, movement-oriented behavior remains largely undertheorized. 
The aim of this study is to bridge that gap: to develop a framework for modeling motivation in specific pedestrian contexts, where internal states are expressed through immediate, observable movement patterns.

In this work, we adapt \textit{expectancy-value theory} \cite{atkinson_introduction_1964, vroom_work_1964, wigfield_expectancy-value_2000, eccles_expectancy_2020} of motivation, originally developed for achievement contexts. Expectancy-value theories attribute motivation to two main components: the \textbf{value} of a potential outcome and the \textbf{expectation} that this outcome can be achieved. 
Both factors are subjective – they are based on the person's own assessment at that moment. 
In achievement context, for example, a person will put effort in studying for an exam if the outcome is highly valued and they believe it is possible to succeed through studying. Conversely, a person who considers passing the exam important but believes success to be unlikely may study less, even if additional studying would objectively improve their chances. 
Over the 60 years of its existence, expectancy-value theory has been refined and extended, with most authors maintaining the core idea that motivation arises from the interaction of value and expectancy. More recent developments, such as situated expectancy-value theory, also takes into account the cultural and milieu-specific contexts under which expectations and values are formed \cite{eccles_motivational_2002}.

The application of expectancy-value theory to pedestrian behavior is, to our knowledge, an innovative approach. While expectancy-value frameworks have typically been used to explain long-term outcomes, their core logic (that motivation is shaped by perceived likelihood of success and the subjective value of the goal) applies well to short-term crowd scenarios. In this context, motivation becomes a key driver of movement behavior, contributing to the emergent dynamics observed at entrances. 

To operationalize this framework for pedestrian dynamics, we first introduce the central concepts of value and expectancy. We then argue that, in this context, expectancy has two components: a spatial component (distance to the goal) and a collective component (payoff structure and relative position within the crowd).  The spatial component is straightforward: as individuals move closer to the goal, their perceived likelihood of success increases. The collective component, by contrast, captures how this likelihood depends on the presence and behavior of others. The payoff structure defines what the relative position to others means in terms of ``winning'' or ``losing.''
It operationalizes expectations in situations with limited resources (such as sequential passage through a bottleneck) under conditions in which many individuals simultaneously attempt to access these resources. In this way, our model explicitly takes the collective context of crowd movement into account.

\textit{Value:} The concept of value refers to the perceived desirability or attractiveness of the potential outcome that can be achieved through individual behavior \cite{vroom_work_1964, porter_managerial_1968}. 
It can be understood as the subjective reward associated with reaching a specific goal. 
This could be the individual significance of a good grade in an exam. In the context of crowds in front of bottlenecks, the value refers to the subjective attractiveness of reaching the bottleneck. The perceived value may vary among pedestrians depending on the urgency of their need to reach the bottleneck. 
For example, some pedestrians may be highly motivated to exit quickly due to time-sensitive obligations (e.g., catching a train or attending an appointment), while others may experience little urgency and assign lower value to early passage. 

\textit{Expectancy:} The concept of expectancy refers to the degree to which a person believes a particular goal is probable \cite{vroom_work_1964}. In simpler terms, it reflects an individual's moment-to-moment assessment regarding the likelihood of achieving a specific goal. The greater the perceived likelihood of success, the higher the motivation of the individual. In our framework, expectancy consists of two components: the spatial position and the collective aspect. The spatial component captures the individual’s position relative to the goal, and the collective component reflects the individual’s position relative to others in combination with the payoff structure. 

Spatial component: In the context of pedestrian dynamics, particularly in bottleneck situations, the initial perceived likelihood of success is primarily determined by an individual’s initial spatial position relative to the bottleneck. Individuals closer to the exit are more likely to believe that success is attainable, while those farther back may anticipate delays. As a person gets closer to the exit, the perceived likelihood of success increases. Accordingly, because positions continuously change, this likelihood also fluctuates with proximity to the exit. 
In this respect, the temporal dimension differs significantly from the original contexts of the theory: in crowds, expectations can change rapidly within seconds, resulting in dynamic changes in motivation. 

Collective component: In addition to the spatial component, crowds are strongly shaped by the presence and behavior of other individuals within the crowd. 
In a situation involving only a single person, movement toward a goal (e.g., an entrance) would mainly depend on that individual's level of motivation. 
However, in crowds, where such scenarios involve multiple individuals, movement is constrained by the physical presence of others. 
Human bodies occupy space, and in bottleneck environments increasing density limits individuals' ability to advance freely. 

Moreover, crowd movement can take the form of a competitive process. Related findings (although not explicitly framed in terms of expectancy-value theory) can be found in the literature on competitive sports. 
For example, studies show that performance and risk-taking change dynamically as interim results become available, with the most motivated individuals turn out to be those who are ranked just behind the leaders \cite{GenakosPagliero2012}. 
Similarly, Lackner \cite{lackner_effort_2023} shows that the presence of a clear ``superstar'' in a tournament can reduce the effort of other participants, as their perceived chances of success decline. 
These observations are consistent with expectancy-value theory: people perceive their own success as less likely as their relative position worsens. 

At the same time, not all crowd situations are structured like a sports competition. In many cases, individuals simply aim to reach a destination collectively.  In evacuations, the objective is not to win but to exit fast enough to not be harmed. 
To capture these different collective contexts, we added a payoff structure to our model. 
Borrowed from game theory (without adopting a comprehensive game-theoretical framework), the payoff structure specifies which outcomes are associated with particular strategies and what possible gains and losses they involve, often represented conceptually as a payoff matrix. 

For individuals in a crowd, the payoff structure determines how their relative position corresponds to potential success or failure. For instance, if participants are instructed to be among the first ten to pass through a bottleneck, being in position fifty would be disadvantageous, even though many others may still be behind them. In this case, only the first ten participants would win and all the others would lose. 
Alternatively, the same group of participants may be instructed to pass through the bottleneck together and succeed only if the overall passage time is as  short as possible, in which case the payoff structure allows all the participants to succeed. Another possibility is a payoff structure in which only a certain proportion of the crowd succeeds. In practice, of course, the payoff structure is not assumed in exact numerical terms but rather through estimated quantities, such as ``only the first few'', ``most of us'', or ``everyone.'' This applies to situations in which the overall number of individuals can be roughly estimated; in larger crowds, however, limited visibility and scale make it difficult to form such estimates, introducing additional uncertainty.

Finally, the dynamic nature described above also applies to payoff. While the payoff structure for a given scenario can be defined in advance, individuals' relative position to others changes continuously. For example, if only the first 10 percent can succeed, being overtaken by others reduces an individual’s likelihood of success.

To sum up, in our framework expectancy contains both a spatial component, such as proximity to the bottleneck, and a collective component, defined by the payoff structure combined with the individual’s relative position to others. It is this interaction between the situation and the person that, we believe, Lewin addressed with his aforementioned definition of motivation. We assume that individuals evaluate their moment to moment situation based on both distance to the bottleneck and their position relative to others. 

To illustrate the interplay of expectancy and value, consider a typical crowd situation. A large number of pedestrians are on the way to a concert. Some indeed arrive early because being close to the performer is particularly important to them, therefore they strongly prefer to be near the stage (value). These individuals might run toward the gate when the venue is opened because they are already close to the entrance and reaching the goal seems plausible (expectancy: spatial expectancy). Others, who share the same value, arrive later (for example because their train was delayed), and find that many people are already queuing. Their expectancy of being among of the first becomes therefore low, so despite their high value they may walk more slowly, as additional effort (e.g., running) would make little difference. 

Ultimately, the payoff structure is determined by the spatial characteristics of the concert venue. How large is it? Are there different areas? What is the total capacity of the venue? In a large concert hall, several hundred visitors may still be able to stand close to the stage, whereas in a small venue only a few dozen people can occupy the front area. This influences how many individuals it is worth trying to be among the first to enter (expectancy: payoff structure).

\subsection{Part Three: Agent based models and motivation} 

This section provides an overview of approaches that investigate the influence of differences in motivation on movement models, as well as on the resulting collective phenomena and transport characteristics. In these approaches, the parameters of the movement models are varied in different ways. In the literature on crowd modeling, the principle of adjusting the parameters of movement models is not exclusively linked to the psychological concept of motivation. Instead, many studies refer to concepts such as emotions, aggression, competition, arousal, or panic. These approaches are included here to provide a comprehensive view of models in which movement parameters are reflected through parameter adaptation. 
To make these different modeling approaches comparable, we introduce an abstract variable $\psi$ that represents psychological variables such as motivation. 
In many models, such a variable is not explicitly defined but is implicitly represented through parameters of movement models. The introduction of $\psi$ therefore serves as a conceptual tool for classification rather than as a directly implemented model variable: most of the works cited below do not define $\psi$ themselves; rather, their parameter usage can be reinterpreted in terms of an implicit $\psi$.
Based on this abstraction, three categories of approaches for modifying behavioral characteristics can be distinguished. 

The first category covers static approaches. 
These models do not represent a psychological variable explicitly; rather, agent parameters are fixed at initialization (uniformly or heterogeneously across agents) and remain constant throughout a simulation. 
In our abstraction, this corresponds to $\psi=\mathrm{const}$. 
Examples include \cite{Helbing2000,Kirchner2003a,Parisi2006,Fischer2020,Rzezonka2022}. 
These types of approaches allow us to systematically examine the effects of differences in collective dynamics, such as those observed between high and low motivation groups.

In the second category, the motion parameters are dynamically adjusted, whereas this adjustment is described by the same state space as the motion itself: $\psi=\psi(t)=f(x_i(t),v_i(t))$. 
This means that motion parameters are calculated based on the current state of the environment, specifically from the positions and velocities of other agents $i$ or from the local geometry of the environment. 
Examples include \cite{Korhonen2011,Moussaid2011,Dietrich2014}. 
This also includes models from robotics, as discussed in detail in the review by \cite{Toll2021}. 
A special case within this class are herd models, in which agents adapt their movement by aligning with the behavior of their neighbors.

In the third category, the motion parameters are also adjusted dynamically, but the agents' state space is extended to include additional variables like $\eta$, representing an arousal or emotional state and $\xi$ representing an externally-specified context of the situation that are independent of the motion state space; see the review \cite{VanHaeringen2022}. 
In these approaches, phenomena such as emotional contagion are of particular interest. 
These typically have the following structure 

\begin{equation*}
\psi(t)=f\bigl(x_i(t),v_i(t),\eta(x,t), \xi, \cdots \bigr).
\end{equation*}

$\eta(x,t)$ is modeled, for example, by using group measures, epidemic dynamics, and contagion processes (see, e.g., \cite{Cao2017b} and the references in \cite{VanHaeringen2022}). 
In most of these cases, however, the changes of $\eta$ are introduced heuristically and lack an explicit psychological foundation (e.g., panic contagion).

In all these categories, the variation in the parameter $\psi$ is primarily used to modify the desired speed $v_0$, the direction of motion $\vec{e}_0$, the magnitude of the speed adaptation, or reaction time $\tau$. 
In some cellular automata models, the strength of interaction or conflict resolution mechanisms are also adjusted in addition to the velocity (see, e.g., \cite{Kirchner2003a}). 
The emphasis on desired speed in many studies can be explained historically by the literature's strong focus on evacuation scenarios and panic modeling. 
In such contexts, panic is often interpreted as an increased motivation to move, which can be directly represented in most movement models as an increase in desired speed. 
This simplified representation is already sufficient to reproduce key phenomena such as clogging at bottlenecks. 
More generally, this approach aligns the structure of many movement models, in which behavioral changes are most easily implemented mathematically through adjustments of the desired velocity vector and, in particular, through the desired speed as part of agents' self-propulsion. 
However, as already outlined in \cite{Durupinar2016} it is more realistic to assume that behavioral changes simultaneously influence multiple movement and interaction parameters.

This discussion highlights three areas of concern: (1) In more complex situations (such as positioning oneself in front of an entrance, on a train platform, or in large public spaces) simply focusing on destination-oriented behavior is often insufficient to describe the behavioral changes. In such situations, other parameters of movement models become more important (e.g., desired distance to others), while motivation continues to play a central role. (2) In many existing modeling approaches, behavioral changes are described using concepts such as panic or emotional contagion. 
To date, however, there is no convincing empirical evidence for a rapid spread of such emotional states in crowds. An interesting finding from the Haghani review (\cite{Haghani2019a}) is that the terms ``panic'' and ``herding'' are primarily used by physicists, mathematicians, and engineers, whereas psychologists and sociologists tend to avoid them in the context of crowd behavior. The papers using the term panic are almost exclusively modeling or conceptual works that provide no empirical evidence for the phenomenon. (3) As outlined above (see Part II \ref{sec:part2}), competitive situations in crowds are usually more complex than simple assumptions suggest. 
In real-world settings, a person's potential success depends not only on their movement speed toward the goal, and the underlying payoff structure is more complex than simple success-or-failure assumptions.

Concretely, the proposed model addresses concerns (1) and (3) above by modulating multiple operational parameters and by representing the payoff structure of the situation; 
concern (2) is sidestepped, since the model does not rely on emotional contagion or on autonomous internal-state dynamics.

The motivation framework proposed in this paper falls within the third category: motivation is a deterministic function of the closeness to achieving 
the goal (SE $\eta$), external situational context (payoff $\xi$),
and of a per-agent value drawn at initialization. 
Three features distinguish it from existing approaches. 
First, the functional form of the modulation is derived from a psychological theory (expectancy-value) rather than chosen heuristically. 
Second, motivation modulates several operational parameters of the movement model simultaneously (desired speed, time gap, neighbor repulsion, interpersonal buffer), not only the desired speed. 
Third, the modulation depends not only on local quantities (distance, neighbors) but also on a global ordering (the agent's rank relative to the whole crowd) through the payoff term, which captures the collective aspect of expectancy.


\section{Model: Extended Value-Payoff (EVP) Model}
\label{sec:model_evp}

The EVP model defines an agent's motivation as a multiplicative combination of value and expectancy, where expectancy is the sum of a spatial and a payoff term.

\begin{equation}
m_i = V_i E_i \quad \text{with} \quad E_i = SE_i + P_i.
\label{eq:motivation}
\end{equation}

First, the motivation $m_i$ of agent $i$ is defined as the product of the agent's value $V_i$ and the overall expectancy $E_i$. The value represents the agent’s intrinsic interest in succeeding, while the expectancy captures how likely the action is to be successful. The expectancy $E_i$ itself is composed of two additive terms: the spatial expectancy $SE_i$ and the payoff $P_i$. Here, $SE_i$ quantifies how likely an action is to succeed based on the distance to the goal, whereas $P_i$ reflects how advantageous the agent’s current relative position is within the crowd. All three quantities depend on the specific setting and the pedestrian's position within it, including the distance to the entrance, the rank within the crowd, and the subjective value attached to the outcome. We now describe the functional form of each term.

\paragraph{Spatial Expectancy.}
Empirical findings from our previous work indicate that behavioral intensity increases as individuals approach a bottleneck; for example, higher levels of pushing behavior are observed closer to the exit \cite{usten_dynamic_2023}. This suggests that the perceived likelihood of success (and thus expectancy) increases as distance to the goal decreases. Accordingly, we model spatial expectancy as a function of an agent’s distance to the exit.

\begin{figure}[htbp]
  \centering
  
  \begin{subfigure}[b]{0.45\textwidth}
    \includegraphics[width=\linewidth]{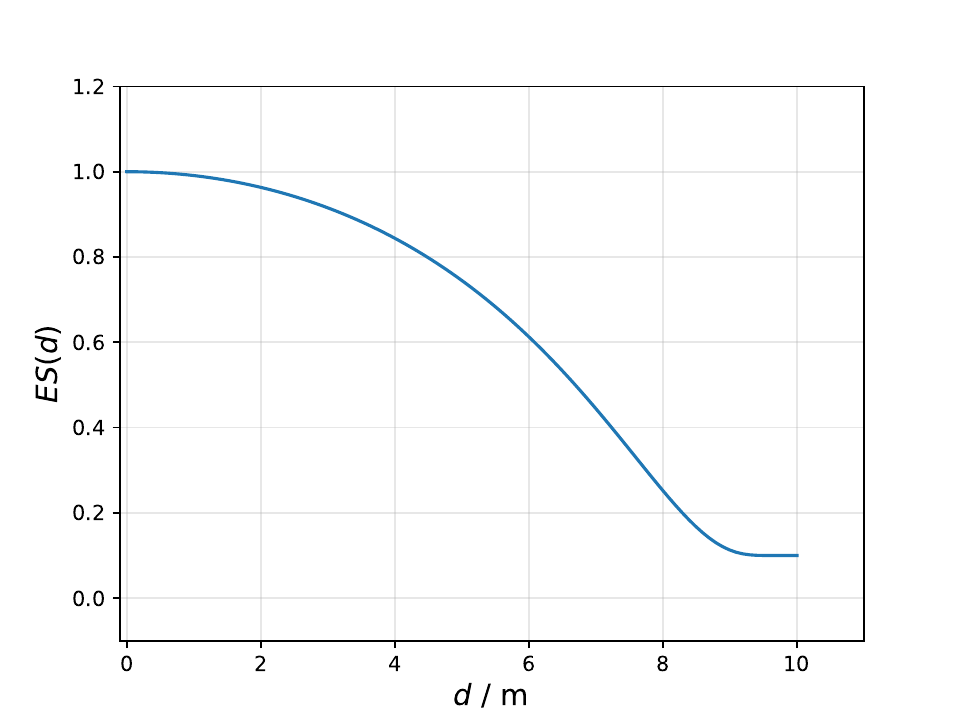}
    \caption{Expectancy $E(d)$ according to \cref{eq:expectancy}.}
    \label{fig:expectancy}
  \end{subfigure}
  \hfill
  \begin{subfigure}[b]{0.45\textwidth}
    \includegraphics[width=\linewidth]{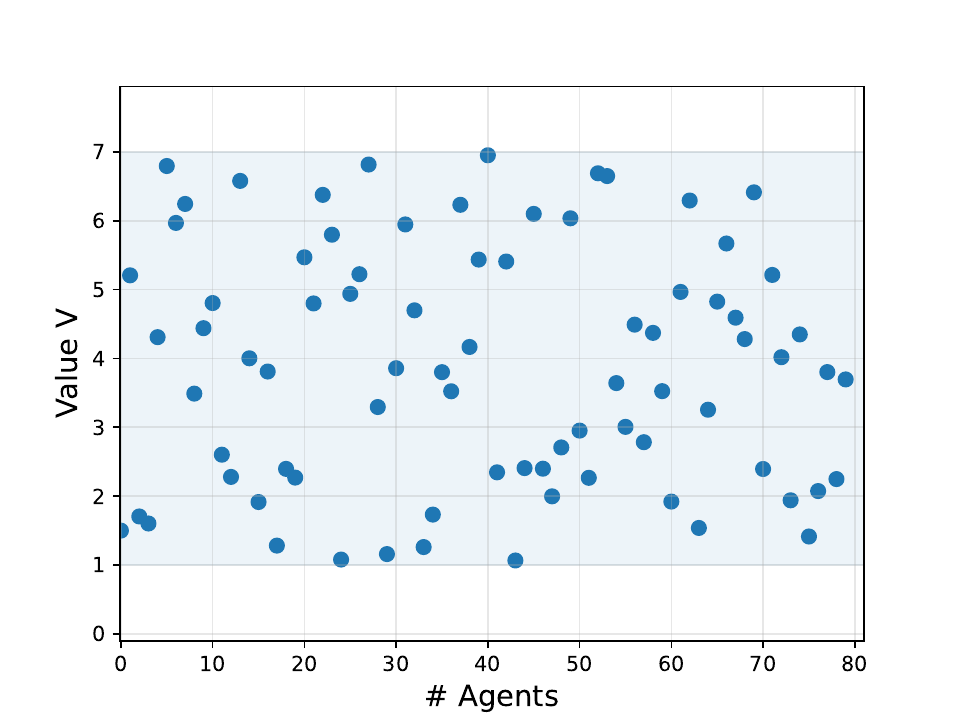}
    \caption{Value assignment $V_i$ for $N=80$ agents drawn uniformly from $[1,7]$ (single seed).}
    \label{fig:value}
  \end{subfigure}
  
  \vspace{0.5cm}
  
  \begin{subfigure}[b]{0.45\textwidth}
    \includegraphics[width=\linewidth]{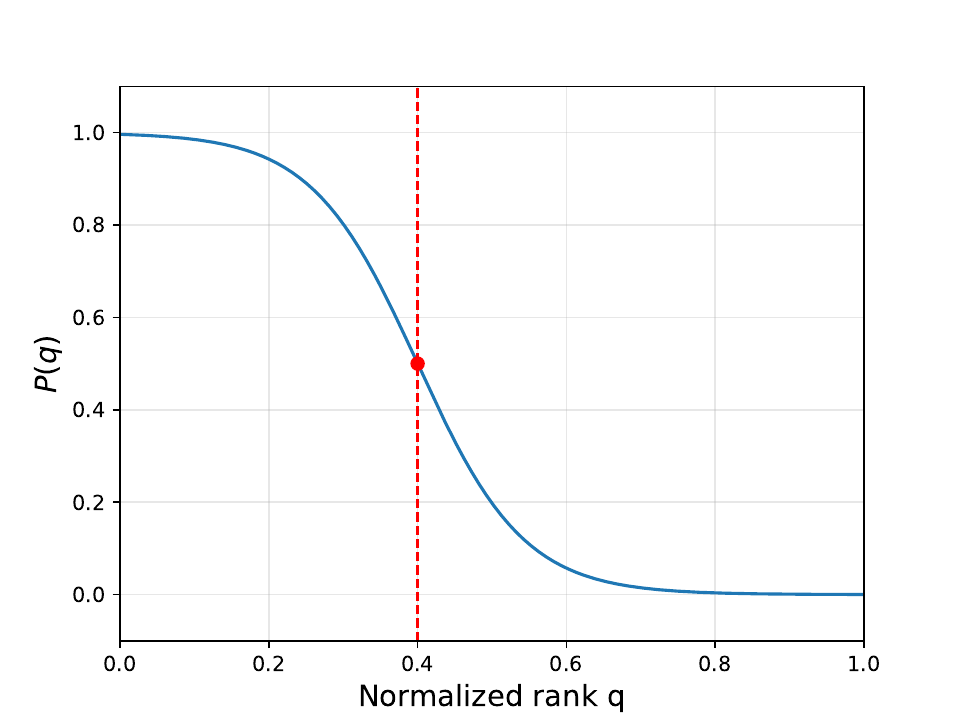}
    \caption{Payoff $P(q)$ according to \cref{eq:payoff}.}
    \label{fig:competition}
  \end{subfigure}
  \hfill
  \begin{subfigure}[b]{0.45\textwidth}
    \includegraphics[width=\linewidth]{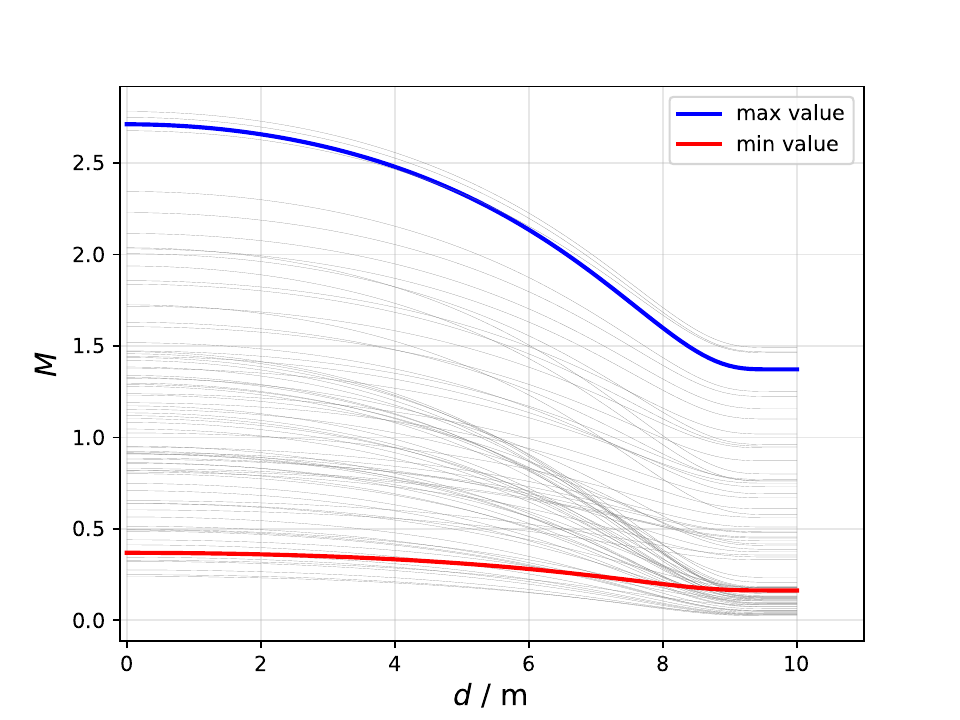}
    \caption{Resulting Motivation according to \cref{eq:motivation}.}
    \label{fig:motivation}
  \end{subfigure}
  
  \caption{Illustration of the EVP components and the resulting motivation (see \cref{eq:motivation}). Each subplot shows how spatial expectancy, value, payoff, and their combination into the overall motivation profile depend on distance, agent identity, or normalized rank.}
  \label{fig:evc_components}
\end{figure}

The spatial expectancy \( SE(d) \) depends on the agent's distance \( d \) to the exit and is defined as

\begin{equation}
SE_i(d) =
\begin{cases}
\varepsilon + (1-\varepsilon)\,e \cdot \exp\!\left(\dfrac{1}{\left(\frac{d}{w}\right)^2 - 1}\right), & d < w, \\[6pt]
\varepsilon, & d \ge w,
\end{cases}
\label{eq:expectancy}
\end{equation}
where \( w \) is the influence radius and $\varepsilon=0.1$ is a baseline floor.
At the goal ($d=0$) the spatial expectancy reaches its maximum value $SE_i=1$; at and beyond $w$ it drops to $SE_i=\varepsilon$, so that distant agents retain a minimal baseline expectancy (Figure~\ref{fig:expectancy}). $e$ is Euler's number.

\paragraph{Payoff.}
The payoff term is based on the current rank of an agent within the crowd.
Let \(r_i\) denote the absolute rank of agent \(i\), obtained by ordering all
agents according to their current distance to the entrance, with \(r_i=1\)
for the agent closest to the entrance. This absolute rank is then normalized to
\(q_i\in[0,1]\):

\begin{equation}
q_i = \frac{r_i - 1}{\max(1, N_{\text{max}}-1)},
\end{equation}
\begin{equation}
P(q_i)= \frac{1}{1+\exp\left(k_P(q_i-q_0)\right)}.
\label{eq:payoff}
\end{equation}
Here, \(r_i\) is the current absolute rank of agent \(i\), \(N_{\text{max}}\) is the scenario maximum reward, \(k_P\) controls the steepness of the payoff decay, and \(q_0\) is the payoff inflection point. 
Agents close to the front therefore receive a higher payoff than agents farther back. The logistic form in Equation~\eqref{eq:payoff} is one convenient choice: it provides a smooth, monotone decrease of payoff with rank, with two interpretable parameters ($k_P$ and $q_0$) controlling steepness and inflection. Other monotone decreasing functions (e.g., linear, exponential, or stepwise reward schedules) are equally compatible with the framework. Figure~\ref{fig:competition} shows the specific parameterization used in the simulations of this paper.

\paragraph{Value.}
Each agent \( i \) is assigned a fixed intrinsic value \(v_i\) at initialization, drawn independently of uniform distribution on $[v_{\min},\ v_{\max}]$.
The raw value is scaled by a fixed constant $\alpha$ before entering the
motivation product:
\begin{equation}
V_i = \frac{v_i}{\alpha},
\end{equation}
where $\alpha = \tfrac{14}{3}$ is chosen so that the maximum possible value
($v_i = v_{\max}$) maps to $V_i = 1.5$, keeping the value
contribution bounded while preserving heterogeneity across agents (Figure~\ref{fig:value}).

\subsection{Motivation in the collision speed model}
Pedestrian motion is simulated in JuPedSim~\cite{jupedsim}  using a collision speed model \cite{Tordeux2016a}; 
The motivational state does not replace this operational model but modulates selected parameters of it.
Since the focus of this paper is the motivation model, we keep the operational description brief here and refer to Appendix~\ref{app:operational-model} for the implementation details.

\subsection{Motivation-to-motion parameter mapping}

Motivation influences the collision speed model by modifying selected operational parameters.

The effective motivation value used for parameter mapping is bounded in order to ensure physically plausible walking speeds.

Operational parameters are then obtained through logistic response
functions,
\begin{equation}
y(m) = y_{\min} + \frac{y_{\max}-y_{\min}}{1+\exp(-k(m-m_0))}.
\label{eq:logistic}    
\end{equation}

Through this mapping, increasing motivation leads to higher desired speeds, reduced temporal gaps, smaller interpersonal buffers, and stronger neighbor repulsion. 
In the present implementation, the neighbor interaction range is also passed through the mapping layer but remains constant at its default value. The detailed anchor values are reported in Appendix~\ref{app:operational-model} and the resulting logistic curves for all five parameters are shown in Figure~\ref{fig:parameter-mappings}.

During each simulation update, the model first evaluates the motivation
level of each pedestrian. The resulting value determines the behavioral
parameters of the operational model, which subsequently governs the
actual motion dynamics.

To ensure that the computed motivation remains within realistic limits, each agent's motivation \( m_i \) is clamped to the interval
\[
m_i \in \left[\varepsilon,\; m_{\max}\right],
\]
where $\varepsilon=0.1$ is the same baseline floor used in the spatial expectancy and $m_{\max}=3.0$ corresponds to the ratio of the maximum physiological walking speed ($v_0^{\max}=3.6\,\text{m/s}$) to the baseline desired speed ($v_0=1.2\,\text{m/s}$).
This motivation value directly influences several behavioral parameters in the model and determines how actively each agent moves within the crowd. 
Agents with higher motivation adjust their behavior in multiple ways. 

First, their desired speed increases proportionally to their motivation, making them more likely to move quickly toward the exit. 
Formally, the adapted desired speed is given by \( \tilde{v}_i^0 = v_i^0 \cdot m_i \). 

In addition, motivated agents close gaps more rapidly because their time gap parameter \( T_i \) shortens with increasing motivation, according to \( \tilde{T}_i = T_i / m_i \).

Motivation also affects how agents interact with others. As motivation rises, both the repulsion strength \( A_i \) and the range \( D_i \) in the direction model increase, which means that motivated agents are more sensitive to their neighbors and more inclined to change direction to overtake and advance through the crowd. 
Finally, the buffer size \( b_i \) is reduced for highly motivated agents, allowing them to keep less distance from their neighbors. 
In contrast, agents with lower motivation tend to maintain larger buffers, showing more passive behavior and being more strongly influenced by surrounding pedestrians.

Together, these adaptations enable the model to reproduce a broad range of realistic behaviors, from passive waiting to more assertive and proactive movement, which may also occur simultaneously within the same crowd. 
Highly motivated agents tend to close gaps more quickly, maintain smaller distances to others, and adjust their direction more actively to advance toward the exit sooner. 
In contrast, agents with lower motivation keep larger gaps, move at lower speeds, and are more likely to follow others without actively seeking better positions. 
\section{Methods}
\label{sec:methods}

\subsection{Experimental reference}
\label{sec:methods-experiment}
The simulations in this work are inspired by the CROMA bottleneck experiments conducted in a multipurpose event hall in D\"usseldorf~\cite{adrian_crowds_2020,sieben_repertoires_2025}. Participants ($85$--$90$ per run) were asked to imagine waiting in front of a concert venue and to enter through a single narrow gate after a short waiting period. Two motivation conditions were realized via the framing of the instruction: in the low-motivation condition (nM) participants were told that their seats were reserved, so there was no need to rush; in the high-motivation condition (hM) no seats were reserved. Throughout this paper we use four runs as the empirical reference, two per condition: \texttt{1C060} and \texttt{2C020} (nM), and \texttt{2C070} and \texttt{2C120} (hM). The trajectory data are publicly available (DOI: \href{https://doi.org/10.34735/ped.2021.14}{10.34735/ped.2021.14}); we process them with PedPy~\cite{schrodter_2025_pedpy} to obtain the same Voronoi-area observable used in the simulations.

The simulation campaign uses $N=80$ agents, slightly below the experimental range, to keep the closed-door waiting geometry comparable while reducing computational cost. The setup is therefore inspired by the CROMA experiment rather than a one-to-one reproduction of it.

\subsection{Simulation setup}
The simulation analysis is restricted to the closed-door setting with two population sizes, $N=40$ and $N=80$. The larger size approximates the CROMA experiment ($85$--$90$ participants per run; see Section~\ref{sec:methods-experiment}). The smaller size $N=40$ is included to inspect the spatial waiting structure at a lower overall density than $N=80$. Open-door scenarios are excluded because the analysis targets the \emph{spatial structure of waiting} -- how agents distribute and pack in front of a closed door -- which is obscured once the door admits throughput. 
In all retained runs the door-opening time ($100\,\mathrm{s}$) exceeds the simulated horizon ($90\,\mathrm{s}$), so the bottleneck remains closed throughout. Across the retained analyses, the comparison is focused on the EVP model and the uniform-motivation baseline, except for the final-rank analysis where all submodels are retained as an ablation of the payoff, value, and expectancy terms. 
The uniform-motivation baseline corresponds to the static baseline in which the motivation-driven parameter modulation is switched off ($m_i \equiv 1$ for all agents), so that all agents share fixed, identical movement parameters.

The simulation campaign uses a paired design with $10$ seeds per model and per population size.
The same seed value is written into the configuration files of every model, so that the initial agent positions and the per-agent value draws are identical across models for each seed. 
This allows paired, within-seed comparisons of the EVP and base outcomes. 
Each per-seed run contributes one value per summary scalar defined below; distributions of these per-seed scalars are compared with a paired Wilcoxon signed-rank test, and Cliff's $\delta$ is reported as effect size. Band plots (median and interquartile range) across the ten seeds replace the earlier single-run illustrations.

The choice of $n=10$ paired seeds per condition follows pragmatic considerations: it is large enough for the paired Wilcoxon test to reach its minimum $p$-value of $0.005$ when all signs agree, and small enough to keep the simulation campaign tractable. 
We do not claim that this value is sufficient for arbitrary observables; convergence-of-variance methods such as the functional-analysis criteria proposed by Ronchi et al.~\cite{Ronchi2014} for ``behavioural uncertainty'' and their multifactor extension by Smedberg et al.~\cite{Smedberg2021} would be the appropriate tools for determining a stricter scenario-dependent run count, and we identify this as a useful direction for follow-up work.

\paragraph{Parameter selection.}
The model's parameters fall into two groups. 
The operational anchors in Table~\ref{tab:motivation_parameters} (desired speed, time gap, buffer, turning bound, neighbor range) are set to physically plausible values consistent with the literature \cite{Tordeux2016a}.
The EVP-specific parameters ($w$, $\varepsilon$, $\alpha$, $k_P$, $q_0$, value range $[v_{\min},v_{\max}]$) are not calibrated against data: at present there is no empirical dataset that links a directly measured motivational state to bottleneck observables, so a fit would be ill-posed. 
We therefore choose these parameters to demonstrate model expressivity rather than to reproduce a specific run. 
The value range $[1,7]$ follows the convention of Likert-type self-report scales and provides a simple ordinal mapping between low- and high-motivation conditions; the contrast in Section~\ref{sec:results} is produced by selecting two sub-ranges of this scale ($[1,3]$ for nM, $[4,7]$ for hM). 
The remaining EVP parameters were chosen by hand to produce non-trivial behavior across the full motivation range and were not tuned per scenario. 
The simulation framework is designed so that any of these parameters can be re-estimated as future empirical work makes calibration possible.

\begin{table}[htbp]
  \centering
  \caption{Motivation model parameters used in the EVP simulation.}
  \begin{tabular}{ll}
    \toprule
    Parameter & Value \\
    \midrule
    Baseline desired speed $v_0$ & 1.2 m/s \\
    Time gap $T$ & 1.0 s \\
    Expectancy width $w$ & 10.0 m\\
    Value range $[v_{\min}, v_{\max}]$ & $[1, 7]$ (uniform) \\
    Value scaling $\alpha$ & 14/3 \\
    Payoff steepness $k_P$ & 14.0 \\
    Payoff inflection $q_0$ & 0.4 \\
    Agent buffer & dynamic (based on motivation) \\
    \bottomrule
  \end{tabular}
  \label{tab:evc_parameters}
\end{table}

\backmatter





\section{Results}
\label{sec:results}
\subsection{Final-rank analysis}
Because the door remains closed in all retained scenarios, the ordering-based analysis is formulated in terms of final rank.
For each agent, the final rank is defined at the last simulation frame by ordering agents according to their distance to the door center, with rank $1$ assigned to the agent closest to the entrance. 
This rank is then related to the agent's mean Voronoi area over the trajectory. 
Rank versus area is used here as a proxy for competitive ordering. 
The final-rank axis is the minimal empirical counterpart of the \emph{collective} component of expectancy introduced in Part~Two: it summarizes where an agent ends up in the payoff ordering relative to others, which is exactly the construct that the payoff term in the model is designed to act on. 
The submodel panels are therefore interpreted as ablations of the model terms, not as additive decompositions of the full EVP response under the nonlinear motivation-to-parameter mapping.

\begin{figure}[htbp]
  \centering
  \begin{subfigure}[b]{0.48\textwidth}
    \includegraphics[width=\linewidth]{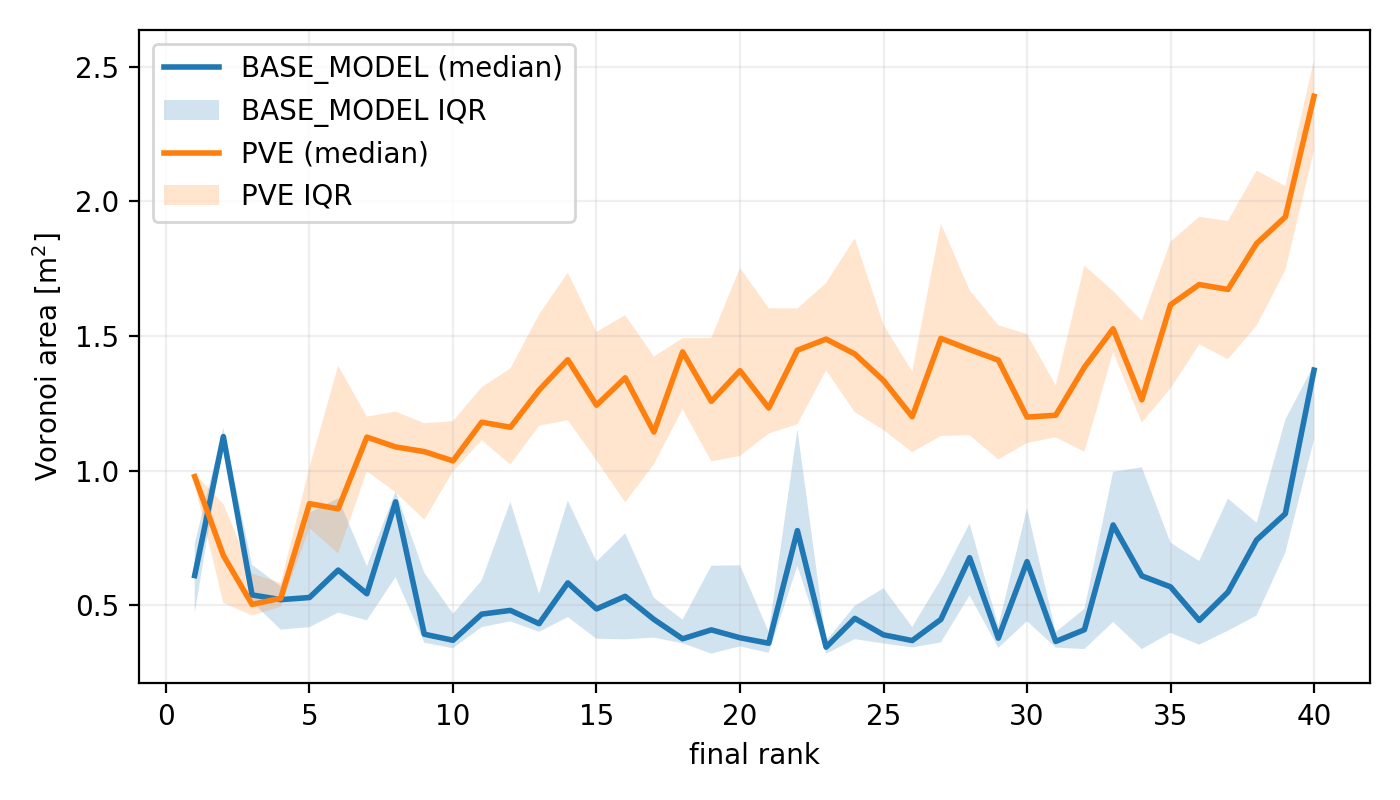}
    \caption{$N=40$, closed door}
  \end{subfigure}
  \hfill
  \begin{subfigure}[b]{0.48\textwidth}
    \includegraphics[width=\linewidth]{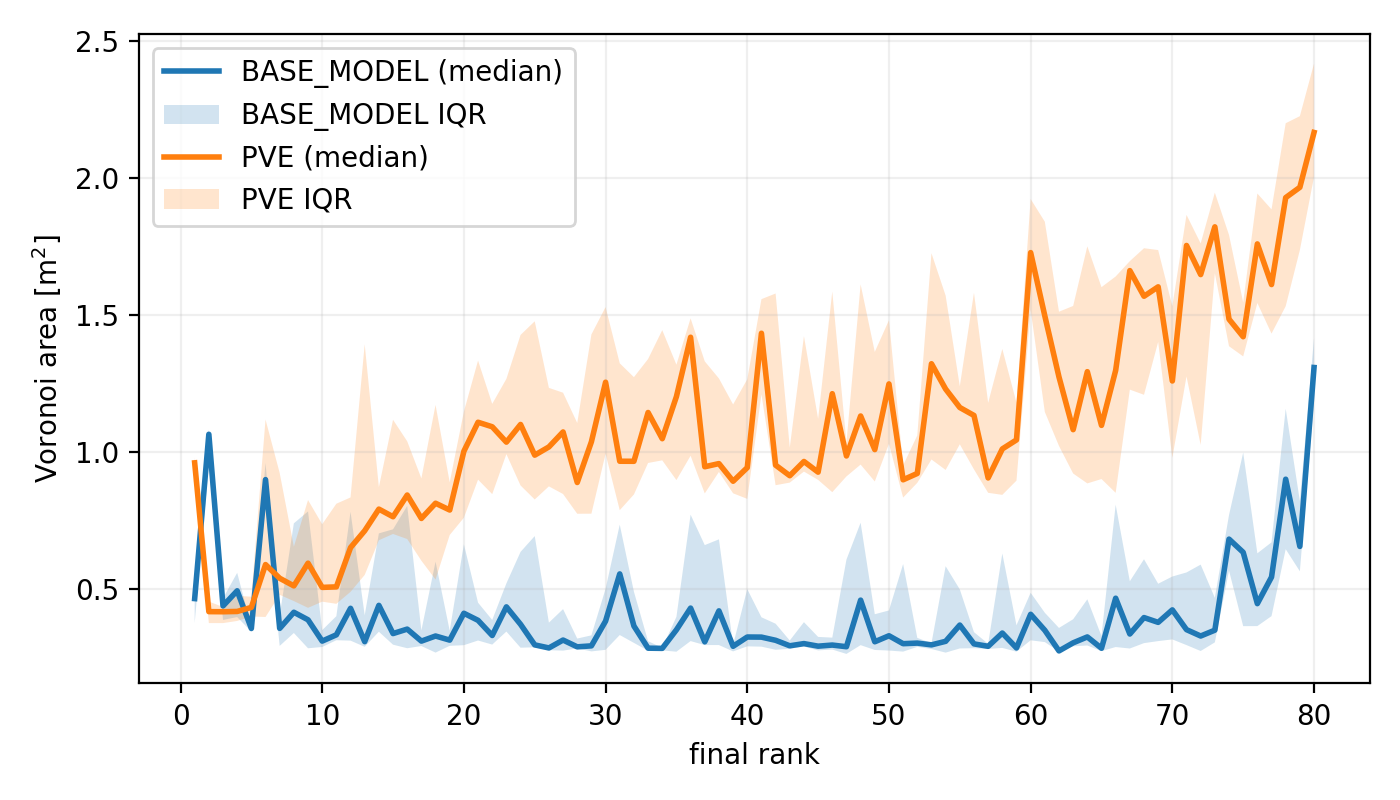}
    \caption{$N=80$, closed door}
  \end{subfigure}
  \caption{Rank--area band plots across ten paired seeds. Solid lines are the across-seed median; shaded regions are the interquartile range. The EVP band rises with rank while the base-model band stays flat, consistent with the statistical tests reported in the text (Wilcoxon $p=0.005$, Cliff's $\delta=1$).}
  \label{fig:rank-area-bands}
\end{figure}

In the retained runs, the uniform-motivation baseline shows a comparatively flat rank--area relation and lacks the upward tail for agents farther back in the crowd. 
The flatness is a property of the rank--area projection: an agent's final rank carries almost no information about how much space it occupied during the run, which is what one expects from a largely homogeneous approach to the door. 
By contrast, the EVP model and, to a lesser extent, the $P$- and $V$-only ablations display a clearer increase of mean area with worsening final rank. 
Across the ten paired seeds, the median Spearman correlation $\rho (\text{rank},\text{area})$ is statistically indistinguishable from zero for the uniform-motivation baseline ($\rho \approx 0.00$ at both $N=40$ and $N=80$) and rises to $\rho \approx 0.58$ at $N=40$ and $\rho \approx 0.66$ at $N=80$ under EVP. 
Paired Wilcoxon signed-rank tests on $\rho$ give $p=0.005$ (the minimum attainable at $n=10$) and Cliff's $\delta = 1.0$ at both population sizes, 
indicating the EVP modulation produces ordering structure absent in the unmodulated baseline. 
The same qualitative pattern holds for the OLS slope and for the top-quartile/bottom-quartile tail ratio. 
We treat Spearman $\rho$ as the primary scalar, with the OLS slope and tail ratio as supporting measures on the same paired data; under Holm--Bonferroni correction across the three tests, all three remain significant at the $n=10$ floor (adjusted thresholds $0.017$, $0.025$, and $0.05$ at $\alpha=0.05$).
We therefore treat the final-rank observable as the clearest quantitative evidence that the motivation model changes ordering structure rather than simply increasing compression everywhere. 
The band plots in Figure~\ref{fig:rank-area-bands} show the median curve and the interquartile range across the ten seeds. 
Per-agent rank--area scatter for the individual $P$-, $V$-, and $SE$-only ablations is shown for reference in Figure~\ref{fig:final-rank-scenarios} in the appendix.
Comparison with alternative motivation models (heterogeneous-$v_0$, local dynamic, contagion-based) is left for future work.

\begin{figure}[htbp]
  \centering
  \begin{subfigure}[b]{0.48\textwidth}
    \includegraphics[width=\linewidth]{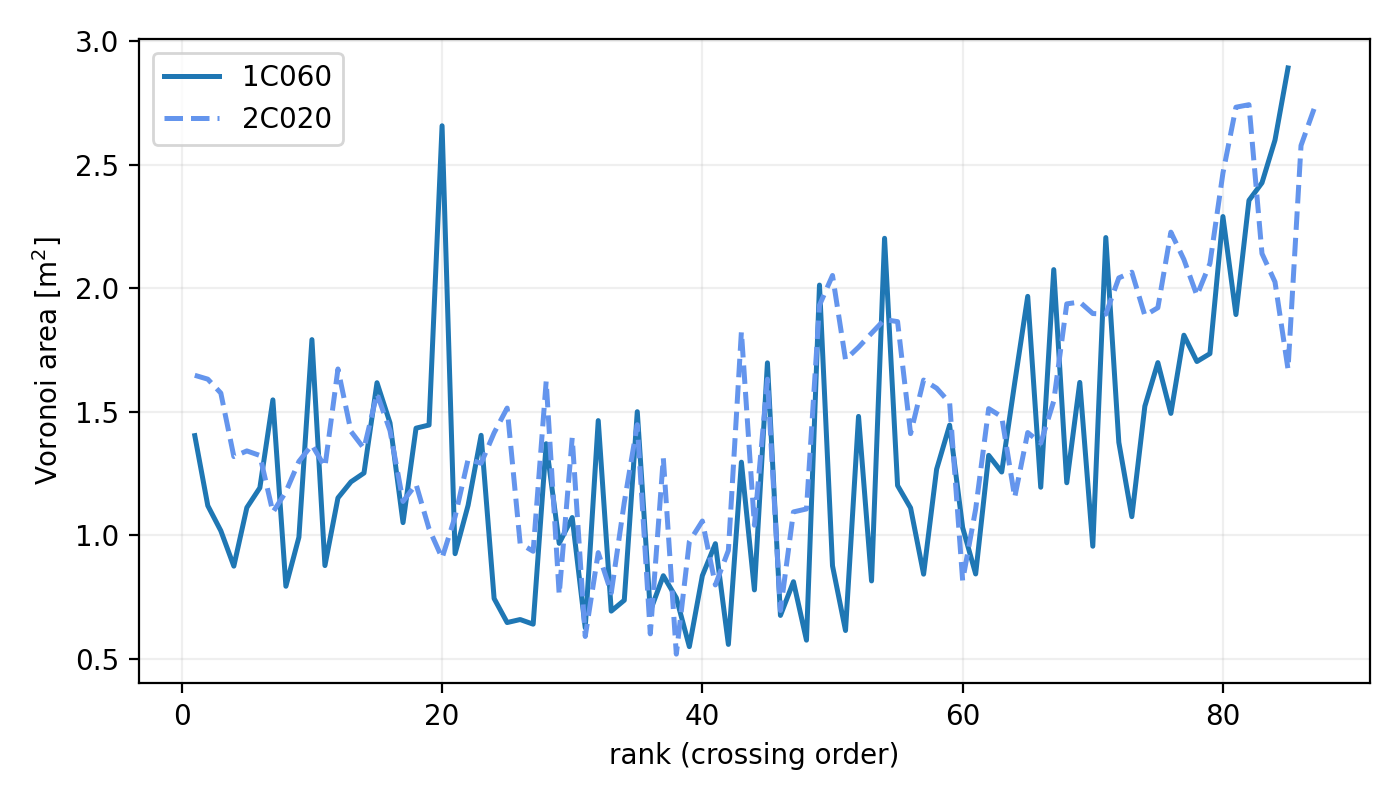}
    \caption{Low motivation (nM)}
    \label{fig:croma-rank-area-nM}
  \end{subfigure}
  \hfill
  \begin{subfigure}[b]{0.48\textwidth}
    \includegraphics[width=\linewidth]{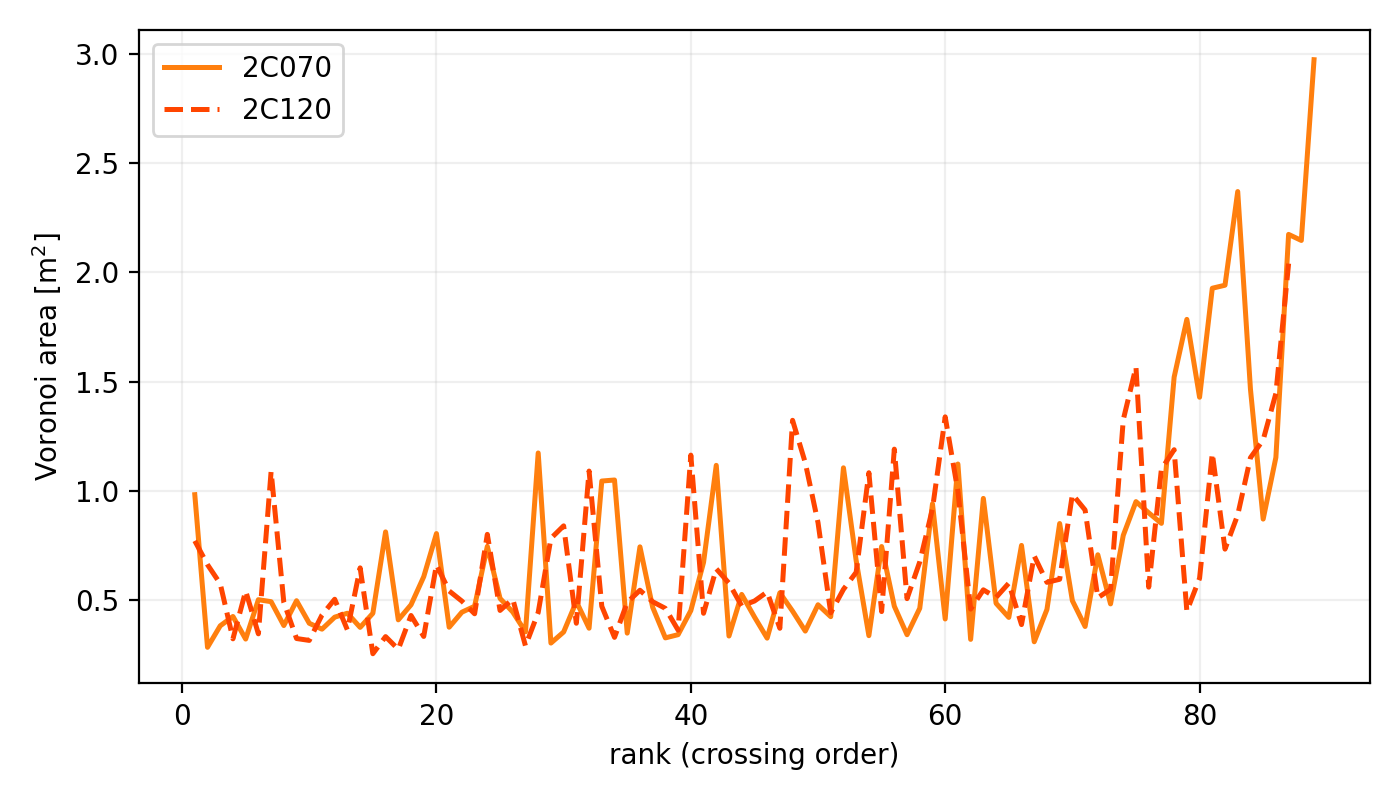}
    \caption{High motivation (hM)}
    \label{fig:croma-rank-area-hM}
  \end{subfigure}
  \caption{Rank--area curves from the CROMA experiments grouped by motivation condition. Rank is defined by crossing order at the door; Voronoi area is computed with the same PedPy settings as the simulation. Both groups show a monotone upward trend (ordering structure is universal), but the hM group operates at a distinctly lower absolute area level (stronger overall compression).}
  \label{fig:croma-rank-area}
\end{figure}

\subsection{Spatial motivation heatmaps}
The spatial motivation analysis is restricted to the closed-door scenarios with $N=40$ and $N=80$. Here the focus is on the EVP model versus the uniform-motivation baseline. All trajectory samples in the interval $10\,\mathrm{s}\le t\le 90\,\mathrm{s}$ are projected onto the $(x,y)$ plane, binned on a $60\times 60$ grid, and smoothed with a Gaussian kernel of width $\sigma=1.5$ bins. The plotted field is the local mean motivation in each spatial bin. Because each panel uses its own color range in the plotting script, the heatmaps are interpreted mainly in terms of spatial localization patterns, not as absolute cross-panel magnitude comparisons.

Under this interpretation, the heatmaps are the spatial counterpart of the final-rank analysis: they visualize the \emph{spatial} component of expectancy introduced in Part~Two, namely proximity to the goal. A concentration of motivation near the advancing front and the bottleneck is the empirical signature expected from the goal-proximity hypothesis~\cite{jhang_impatience_2015}, and it shows that the EVP state is not simply a uniform increase in urgency across the whole crowd. 
Motivation remains spatially structured, which is consistent with the model producing differentiated waiting behavior rather than a globally intensified rush.

\begin{figure}[htbp]
  \centering
  \includegraphics[width=\textwidth]{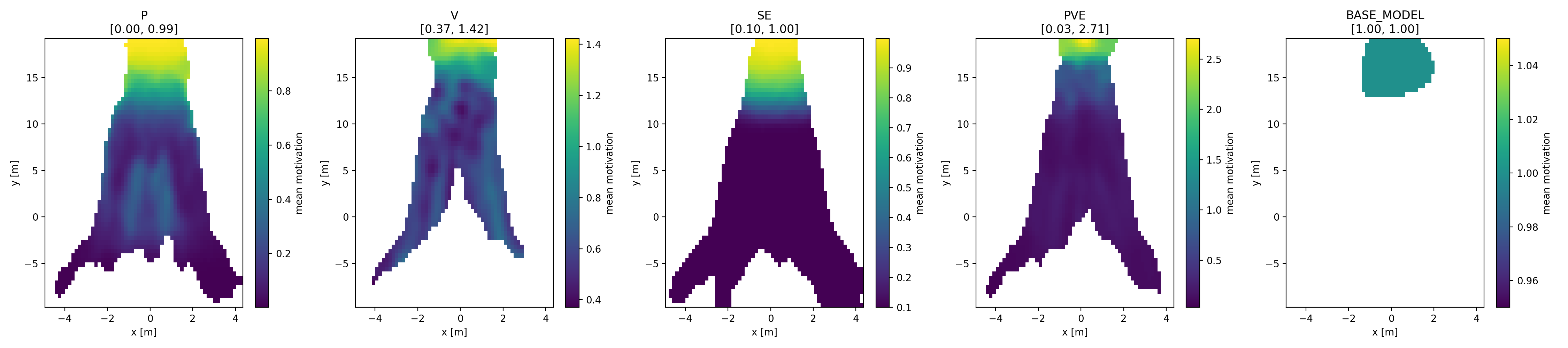}
  \caption{Spatial heatmap of the local mean motivation for $N=80$ with a closed door. The $N=40$ case is qualitatively similar and is omitted for space. Each panel shows one submodel; motivation remains spatially concentrated near the advancing front and the bottleneck, consistent with the spatial component of expectancy defined in Part~Two.}
  \label{fig:motivation-heatmaps-scenarios}
\end{figure}

\subsection{Trajectories}
The spatial structure of the simulated trajectories makes the differentiation visible directly. Figure~\ref{fig:sim-trajectories} shows trajectory traces for a single seed ($N=80$, closed door), colored by each agent's mean motivation. 
In the uniform-motivation baseline all agents share the same motivation ($m=1$) and spread broadly through the corridor. 
In the $SE$-only run a strong distance gradient appears (high motivation near the door, low far away), while $V$-only shows agent-level color differences but a similar spatial spread. The $P$-only run introduces rank-dependent variation. 
The combined $EVP$ model produces the most structured picture: high-motivation agents (yellow) cluster tightly near the bottleneck, while low-motivation agents (purple) remain further back with wider spacing.

\begin{figure}[htbp]
  \centering
  \includegraphics[width=\textwidth]{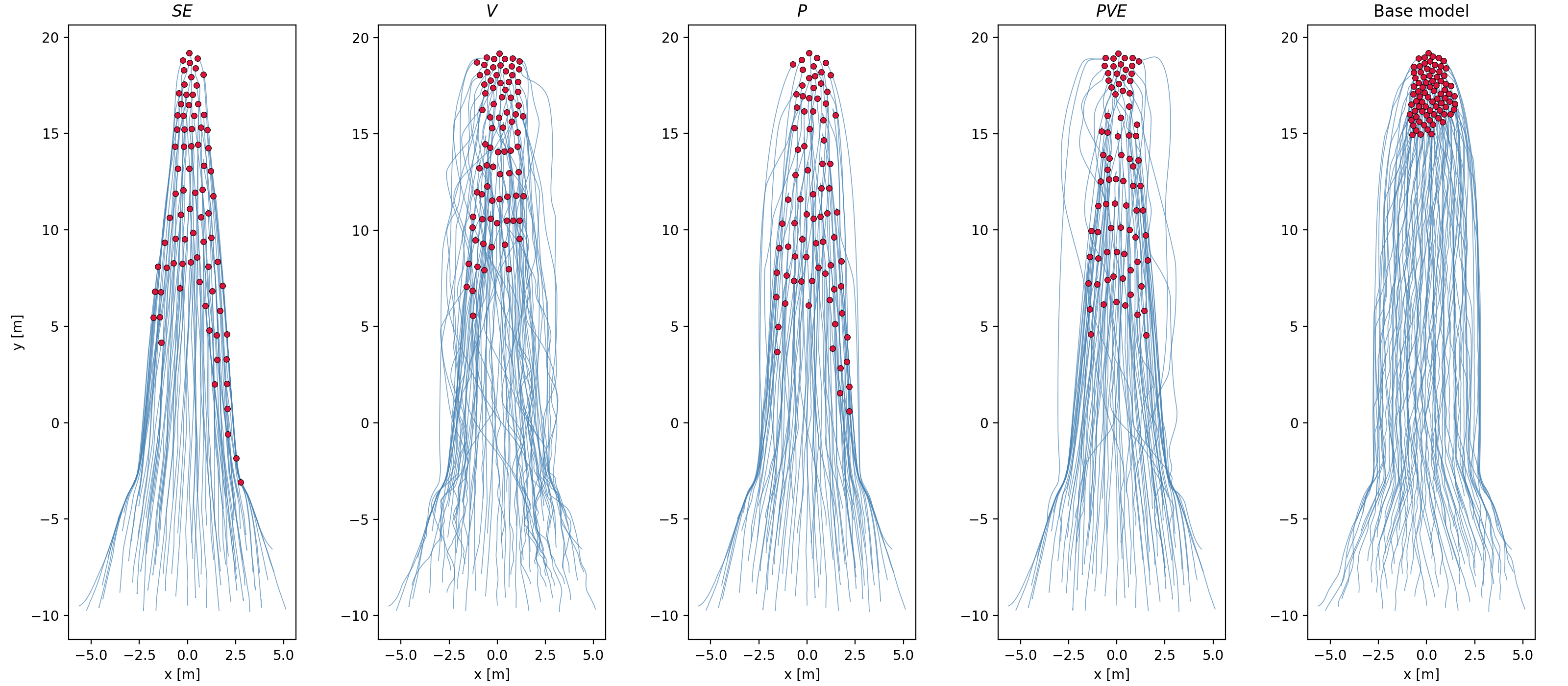}
  \caption{Simulated trajectories ($N=80$, closed door, single seed) for each submodel, plotted in a single color for direct visual comparability with the CROMA experimental trajectories in Figure~\ref{fig:croma-trajectories}. Filled markers indicate each agent's final position.}
  \label{fig:sim-trajectories-uniform}
\end{figure}

\begin{figure}[htbp]
  \centering
  \includegraphics[width=\textwidth]{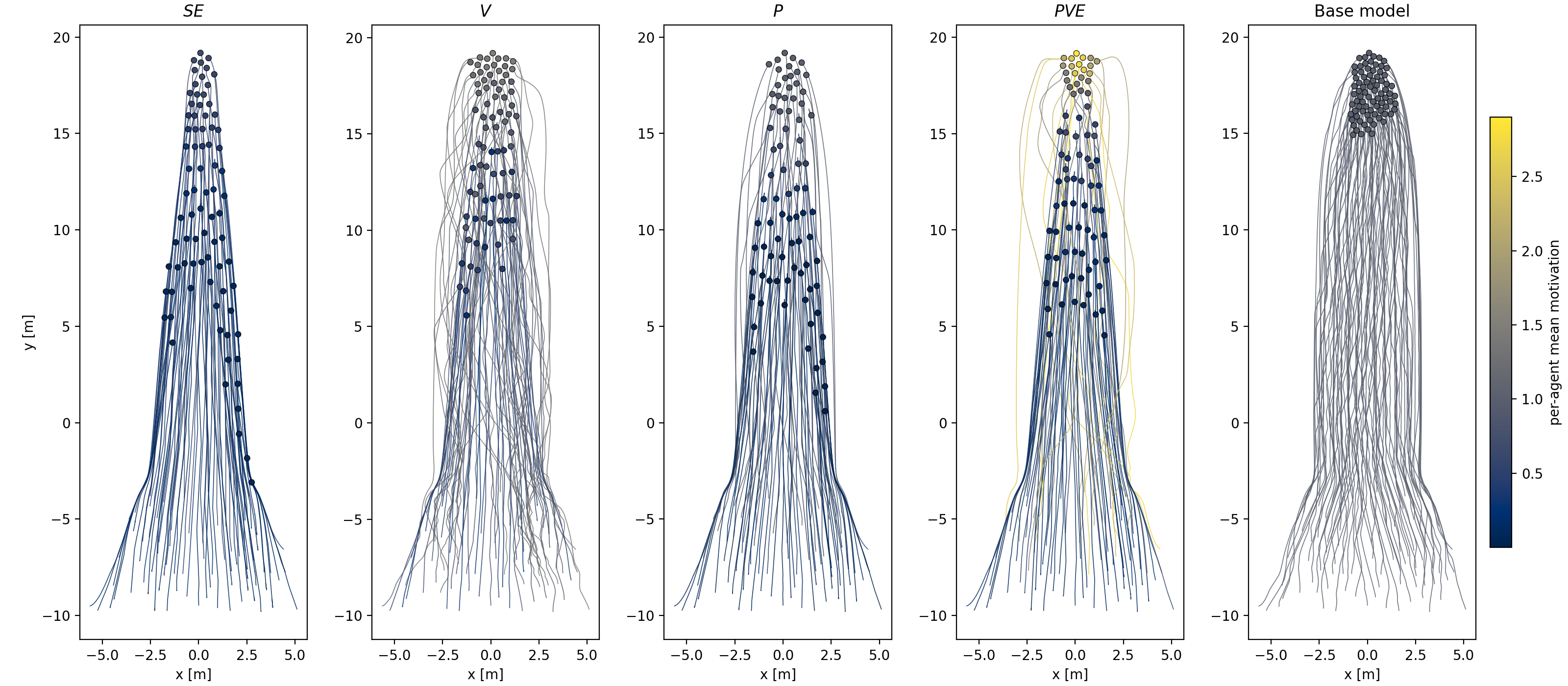}
  \caption{Same simulated trajectories as in Figure~\ref{fig:sim-trajectories-uniform}, now colored by each agent's mean motivation (cividis colormap; darker blue $=$ lower, lighter yellow $=$ higher). The uniform-motivation baseline produces uniform, broadly spread trajectories. The individual terms $P$, $V$, and $SE$ each introduce partial differentiation. The combined $EVP$ model yields the most structured pattern, with high-motivation agents converging tightly near the bottleneck.}
  \label{fig:sim-trajectories}
\end{figure}

\begin{figure}[htbp]
  \centering
  \includegraphics[width=\textwidth]{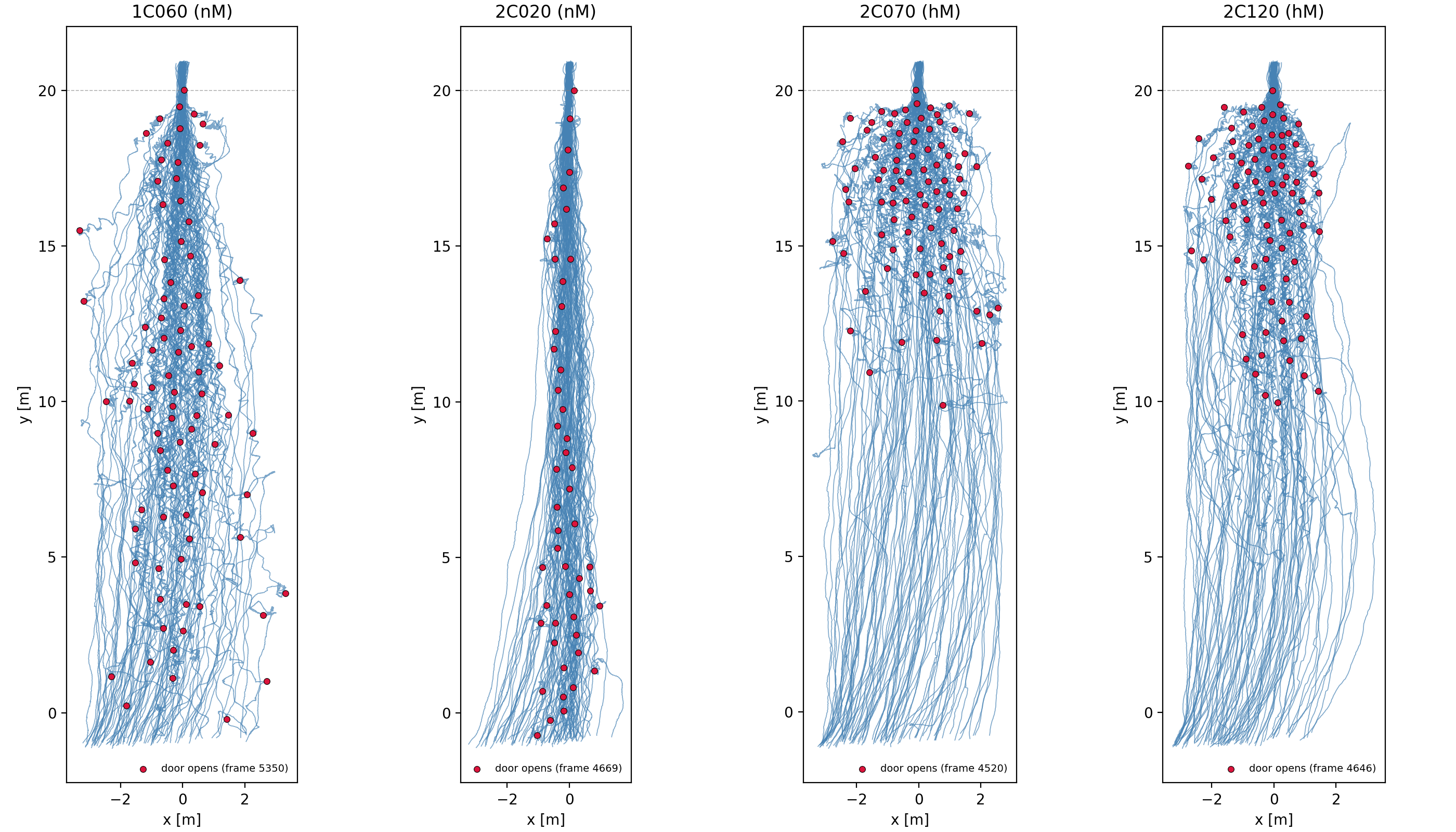}
  \caption{Trajectories of participants for two low-motivation runs (nM: \texttt{1C060}, \texttt{2C020}) and two high-motivation runs (hM: \texttt{2C070}, \texttt{2C120}) from the CROMA experiments~\cite{adrian_crowds_2020,sieben_repertoires_2025}.
  Trajectories show that participants moved from the lower part through the corridor toward the bottleneck.
  Knots in the trajectories indicate where participants waited for the gate to open.
  Filled markers show each agent's position at the moment the door opens; the dashed line at $y=20\,\mathrm{m}$ marks the door line.
  The nM runs show wider lateral spread, while the hM runs converge more directly toward the entrance.}
  \label{fig:croma-trajectories}
\end{figure}

\begin{figure}[htbp]
  \centering
  \includegraphics[width=\textwidth]{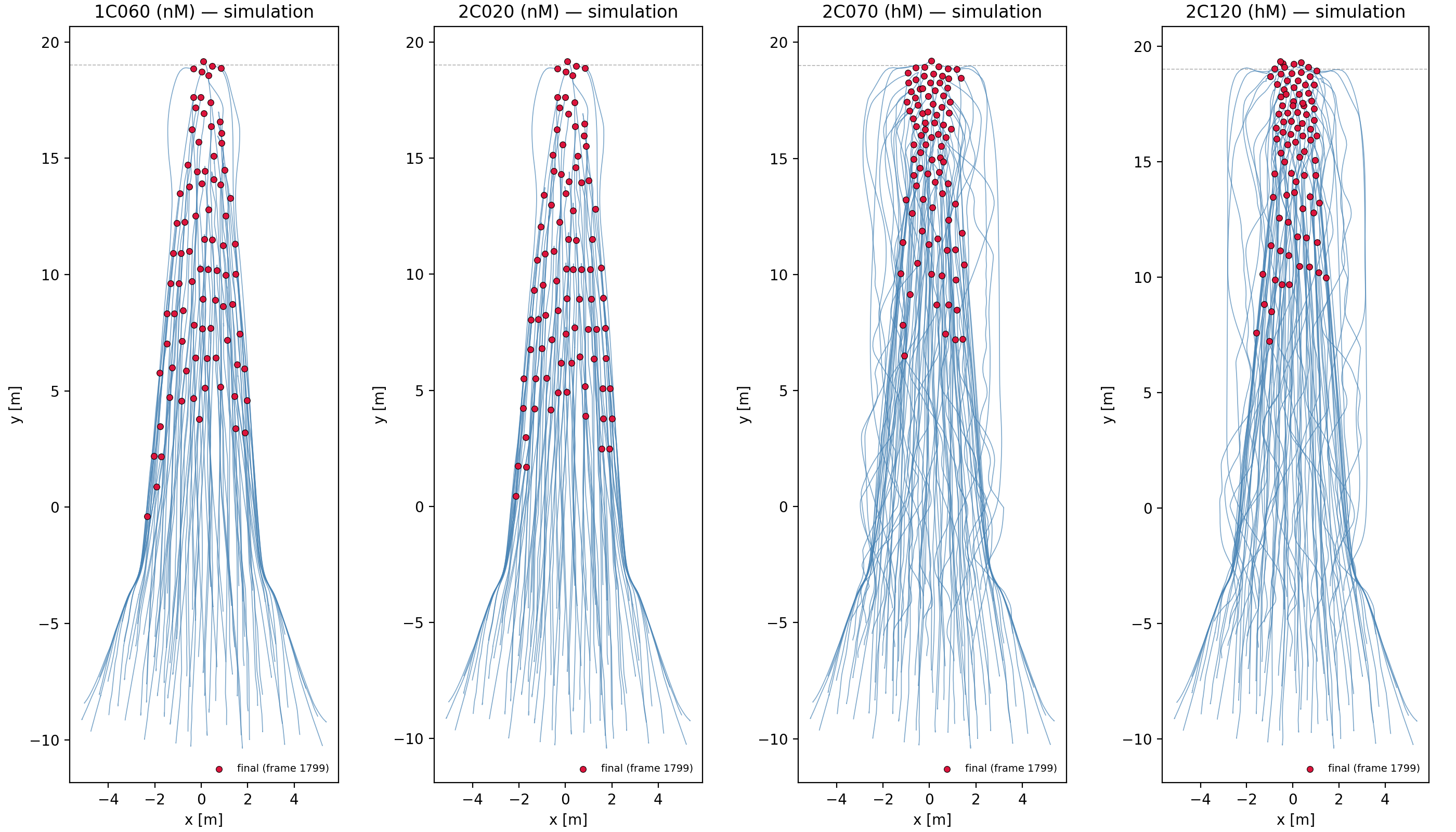}
  \caption{EVP simulations matched one-to-one to the four CROMA runs in Figure~\ref{fig:croma-trajectories}. Each panel uses one configuration parameter per condition: the value range $[v_{\min}, v_{\max}]$ is set to $[1,3]$ for the two nM panels (low intrinsic value, hence low motivation) and to $[4,7]$ for the two hM panels (high intrinsic value, hence high motivation). All other parameters are identical across panels: EVP motivation mode, closed door (open-time $> $ horizon), $N$ matched to the experiment ($85$, $87$, $89$, $87$), and the same EVP/payoff parameters as in Table~\ref{tab:evc_parameters}. Agents are placed by random distribution in the simulation corridor (not at experimental positions); the comparison is therefore on \emph{trajectory shape}, not on initial coordinates. Filled markers show each agent's final position; the dashed line at $y=19\,\mathrm{m}$ is the simulation door line.}
  \label{fig:sim-testing-trajectories}
\end{figure}

The simulation reproduces the qualitative direction of the experimental contrast: the nM panels in Figure~\ref{fig:sim-testing-trajectories} produce a comparatively narrow funnel toward the door, while the hM panels develop a wider, denser cluster of agents near the entrance, mirroring the basic distinction between Figure~\ref{fig:croma-trajectories}'s nM and hM runs. The agreement is, however, only qualitative. The lateral spread in the simulated hM panels is visibly smaller than in the experiments, and the simulation lacks the pronounced ``waiting knots'' visible at mid-corridor in the experimental nM runs. This is partly a geometric mismatch: the simulation uses a narrow corridor as its waiting area, whereas the CROMA participants waited in a much larger hall behind the entrance, which mechanically allows wider lateral spread. We therefore read this experiment as evidence that the value range alone is sufficient to shift the spatial signature in the predicted direction, but that fully reproducing the experimental shape would require a closer geometric match to the CROMA waiting hall and is left for future work.

A broader comparison with the CROMA bottleneck experiments~\cite{adrian_crowds_2020,sieben_repertoires_2025} places the simulation claims in empirical context. The four experimental runs are grouped into two motivation conditions: low motivation (nM: \texttt{1C060} and \texttt{2C020}) and high motivation (hM: \texttt{2C070} and \texttt{2C120}). The trajectory overview in Figure~\ref{fig:croma-trajectories} shows the raw spatial spread: nM trajectories fan out more widely and exhibit visible waiting knots at mid-corridor, whereas hM trajectories converge more narrowly and directly toward the bottleneck.

The rank--area analysis (Figures~\ref{fig:croma-rank-area-nM} and~\ref{fig:croma-rank-area-hM}) uses crossing order at the door as the experimental counterpart of the simulation's final-rank ordering, with the same PedPy~\cite{schrodter_2025_pedpy} Voronoi-area observable. Four observations emerge.

First, all four experimental scenarios show a clearly monotone rank--area pattern (Spearman $\rho$ between $0.39$ and $0.59$), so the flat, rank-independent response of the uniform-motivation baseline is empirically ruled out: real crowds are never as homogeneous as the uniform-CSM baseline.

Second, the experimental correlations sit \emph{between} the base and $EVP$ simulations ($\rho\approx 0$ and $\rho\approx 0.58$--$0.66$ respectively). Notably, the simulated $EVP$ range now overlaps with the experimental range, indicating that the current parameterization produces a rank--area effect of realistic magnitude.

Third, the rank--area \emph{correlation} is nearly identical between the two motivation groups (nM mean $\rho=0.49$, hM mean $\rho=0.51$). The ordering structure is therefore universal across the experiments and does not depend on whether participants received a low- or high-motivation instruction. This is consistent with the model's architecture: the expectancy and payoff terms ($SE+P$) produce the ordering, and they act on spatial position and rank, both of which are present regardless of motivational context.

Fourth, the two groups differ markedly in \emph{absolute area level}: hM agents occupy systematically smaller Voronoi cells across all ranks (Figures~\ref{fig:croma-rank-area-nM} and~\ref{fig:croma-rank-area-hM}). This maps onto the value term ($V$) in the model, which scales the overall motivation and therefore the compression. 
The CROMA observations are consistent with a model in which \textit{SE+P} generates the ordering structure and $V$ scales overall compression. 
With four runs we cannot test this decomposition statistically; it remains a model-side prediction.


\section{Discussion}
\label{sec:discussion}

The closed-door scenario analysis suggests that the main contribution of the EVP model is not a simple increase in compression or local density. Instead, it produces more differentiated (both moving toward and waiting) behaviors near the bottleneck. The evidence rests on the three observables that correspond to the components of the underlying expectancy--value construct (Part~Two): the \emph{collective} component (relative position and payoff structure) of expectancy, the \emph{spatial} component (proximity to the goal) of expectancy, and value. Along the collective component, the final-rank versus area plots show that the EVP runs develop a pronounced rank-dependent structure, while the uniform-motivation baseline remains comparatively flat, consistent with a more homogeneous approach to the door. 
Along the spatial component, the motivation heatmaps show that motivation remains localized near the advancing front and the bottleneck rather than rising uniformly across the crowd. 
Value, on the other hand, introduces distributed differences across agents, creating a distinct heterogeneity even within the structured ordering by rank and spatial position, which can be consistently observed across all analyses and figures.
Taken together, these three views describe the EVP model as a mechanism for structured competitive ordering -- in rank and in space -- combined with persistent individual heterogeneity, rather than as a mechanism for uniformly increasing congestion. In this sense, the base-versus-EVP contrast is the empirical manifestation, at the level of a single closed-door bottleneck, of the static-to-dynamic transition advocated in Parts~One and~Three: where the static baseline produces a largely homogeneous crowd with a weak rank signature, the dynamic EVP formulation produces the structured heterogeneity that experimental studies on motivated crowds have repeatedly reported~\cite{adrian_crowds_2020,usten_pushing_2022,usten_dynamic_2023}.

A key implication of these results is that motivational dynamics operate on short time scales (seconds), in contrast to traditional applications of expectancy–value theory. In this work, the theoretical assumptions were adapted to this temporal scale, and operationalized and tested within a simulation framework. The resulting dynamics are consistent with the initial hypotheses and, importantly, with empirical observations: moment-to-moment changes in expectancy and value translate into observable behavioral changes and, at the collective level, into structured crowd patterns observed in real-world settings.

\subsection{Limitations}
The present scenario analysis is intentionally narrow. It uses ten paired seeds per condition, which is sufficient for the paired Wilcoxon tests and effect-size reporting used here, but small for more sensitive inference; a larger sweep of twenty or more seeds would tighten the confidence intervals on the band plots and allow detection of weaker effects. 
The analysis is restricted to a single bottleneck geometry and one population size. However, the underlying mechanism is not specific to this setting: whenever movement is constrained by spatial structure and individuals evaluate their position relative to others under a given payoff structure (such as lane merging, or crowding in open spaces near points of interest), similar patterns of structured heterogeneity and dynamic changes are expected to emerge.
In addition, the current model does not yet include social-group cohesion (but see for an overview of social groups in pedestrian models \cite{Templeton2015}), explicit norm-following behavior, or richer behavioral repertoires \cite{sieben_repertoires_2025}. 
These factors are visible in the experimental material that we used for comparison and they may alter both local density and ordering patterns.
However, the payoff structure can also be interpreted as a simplified proxy for these factors, as it captures how situational constraints and implicitly shared expectations or norms translate relative position into advantages or disadvantages, even though the underlying social processes themselves are not explicitly modeled.
Finally, the EVP parameters are not empirically calibrated; in the absence of data linking a directly measured motivational state to crowd observables, the parameter values reported here serve to demonstrate model expressivity, not to provide a fitted match to a specific scenario.

\subsection{Conclusion}\label{conclusion}

In the retained closed-door scenarios, the EVP-based motivation model is most informative when the analysis focuses on ordering structure rather than on density alone. The clearest qualitative result is that the uniform-motivation baseline produces a comparatively flat rank--area relation, whereas the EVP model generates a more differentiated pattern that is more consistent with heterogeneous waiting behavior near the bottleneck.

The main conclusion is therefore that motivation in this setting should be understood less as a global increase in urgency and more as a mechanism that reorganizes competition along the two axes that the expectancy component of the model predicts: \emph{space} (proximity to the goal) and \emph{rank} (relative position in the payoff structure). This provides a clearer behavioral rationale for the model and helps explain why density-based observables alone are insufficient to evaluate it.

\bmhead{Acknowledgements}

The authors acknowledge the use of AI-assisted tools (ChatGPT, Claude, DeepL Translator) for language editing and code development support. All scientific ideas, analyses, and interpretations are solely those of the authors.

\section*{Funding}
The authors ratefully acknowledges financial support from the BMFTR project Croma-Pro, grant number DB002398 and from the EU under program: Horizon, topic: ERC-2025-SyG, Project ID: 101225087, Project CrowdING.

\section*{Declaration of competing interest}
The authors declare that they have no known competing financial interests or personal relationships that could have appeared to influence the work reported in this paper.

\section*{Data availability}
Experimental data used for comparison are publicly available under \url{https://doi.org/10.34735/ped.2021.14}.
Simulation results, analysis outputs, and all figures presented in this paper are archived on Zenodo~\cite{chraibi_2026_19448415}.

\section*{Code availability}
The source code for the EVP motivation model and all analysis scripts is openly available on GitHub at \url{https://github.com/PedestrianDynamics/Motivation}~\cite{chraibi_2026_19448415}, released under the MIT license.
An interactive interface for data exploration is available at \url{https://pedestriandynamics-motivation.streamlit.app}.

\section*{Author contribution}
Ezel Üsten: Conceptualization, Writing – original draft, Writing – review \& editing. 
Anna Sieben: Conceptualization, Writing – review \& editing, Funding acquisition, Resources. 
Mohcine Chraibi: Writing – review \& editing, Software, Methodology, Formal Analysis, Validation. 
Armin Seyfried: Conceptualization, Writing – review \& editing. 

\section*{Materials availability}
Not applicable.
\section*{Ethics approval and consent to participate}
Not applicable.
\section*{Consent for publication}
All authors consent for publication.

\begin{appendices}

\section{Operational model details}\label{app:operational-model}

Pedestrian motion is described by an operational model built upon the
Collision Free Speed Model (CSM) \cite{Tordeux2016a}. While the
spacing--speed relation of the original model is retained as the conceptual
basis, both the steering dynamics and the effective spacing formulation are
restructured. In particular, the present formulation introduces a rotational
direction model together with an explicit decoupling of speed and direction.
These modifications preserve the conceptual advantages of the CSM while
avoiding common artifacts such as oscillatory zigzag motion or premature
deadlocks at moderate densities.

In addition to the modified steering dynamics, the spacing formulation used
in the speed function is extended by introducing a motivationally dependent buffer term. The parameter $l$ defines the minimum interpersonal distance
between two agents. Its smallest possible value is determined by physical
body size, but social factors also influence this distance. To incorporate
motivational effects into the operational dynamics, an additional buffer
term $b$ is introduced. This term reflects the observation that highly
motivated individuals tend to accept smaller interpersonal distances than
less motivated ones. The effective spacing used in the speed function is
therefore defined relative to the combined distance threshold $l + b$.

\subsection{Decoupling of direction and speed}

In the original CSM formulation, the walking direction results from the
normalized sum of several influence vectors, including the desired
direction and repulsive contributions from neighboring pedestrians and
boundaries. The resulting direction vector is then used to evaluate the
available spacing in front of the pedestrian, which determines the walking
speed through the optimal velocity relation.

This formulation implicitly couples steering and speed selection. The
magnitude of the summed influence vector affects the normalization step and
therefore the direction along which spacing is evaluated. Variations in the
steering field may therefore indirectly influence the perceived available
space and consequently alter the walking speed.

In the present formulation, direction and speed are explicitly decoupled.
The velocity of pedestrian $i$ is written as

\begin{equation}
\dot{\mathbf{x}}_i = V_i(s_i) \cdot \mathbf{e}_i,
\end{equation}
where $\mathbf{e}_i$ is the walking direction determined by the direction
model and $V_i(s_i)$ is the speed obtained from the spacing--speed relation
of the CSM.

\subsection{Rotational direction model}

Let $\mathbf{p}_i$ denote the position of pedestrian $i$ and
$\mathbf{p}_j$ the position of neighbor $j$. The relative position vector
is $\mathbf{r}_j = \mathbf{p}_j - \mathbf{p}_i$. Let
$\mathbf{e}_{\mathrm{des}}$ denote the desired direction and
$\mathbf{e}_{\mathrm{ref}}$ the reference direction after wall
corrections. 
For each forward neighbor
satisfying $x_j = \mathbf{e}_{\mathrm{ref}}\cdot\mathbf{r}_j > 0$, the
signed lateral position is computed using the two-dimensional cross product
$s_j = \mathbf{e}_{\mathrm{ref}} \times \mathbf{r}_j$.

Neighbor relevance is determined through the anisotropic weighting
\begin{equation}
w_j =
\exp\!\left(-\frac{x_j}{D\cdot r_x}\right)\cdot
\exp\!\left(-\left(\frac{y_j}{D\cdot r_y}\right)^2\right),
\label{eq:weighting}
\end{equation}
where $y_j = |s_j|$ denotes the lateral distance from the center line. The
constants in \cref{eq:weighting} are such that $r_x>r_y$.

To maintain a minimal interaction structure, only the most relevant
forward neighbor is considered, $j^* = \arg\max_{x_j>0} w_j$. The
resulting turning angle is
\begin{equation}
\theta = \theta_{\max}\tanh(a_{j^*}),
\label{eq:theta}
\end{equation}
with $a_j = -\, w_j \, \frac{s_j}{|s_j|+\varepsilon_s}$.
In the implementation, the effective turning bound $\theta_{\max}$ is set
by the motivation-dependent neighbor-repulsion strength and clipped by the
global upper bound of the operational model.

The walking direction is then obtained by rotating the reference
direction,
\begin{equation}
\mathbf{e}_{\mathrm{move}} = R(\theta)\,\mathbf{e}_{\mathrm{ref}}.
\label{eq:rotation}
\end{equation}
Because the update is purely rotational, the direction vector remains unit
length by construction.

\subsection{Spacing and speed}

We introduce a blended desired direction
\begin{equation}
\mathbf{s}=(1-w_b)\,\mathbf{s}_{\mathrm{move}}+w_b\,\mathbf{s}_{\mathrm{goal}},
\end{equation}
with a small blending weight $w_b=0.15$, to explicitly balance local
collision-free motion and global progress toward the exit. Here,
\(\mathbf{s}_{\mathrm{move}}\) captures short-horizon, interaction-aware
navigation, while \(\mathbf{s}_{\mathrm{goal}}\) stabilizes long-range
intent and keeps agents aligned with evacuation objectives. This blending
is motivated by the observation that purely local steering can overreact
in dense bottlenecks, amplifying lateral conflicts and stop-and-go
patterns.

In the original CSM, only a single direction \(\mathbf{s}\) is used. In
practice, this can favor clogging, because directional updates are
dominated by immediate local constraints without a compensating
goal-oriented term to restore coherent outflow.

The walking speed follows the optimal velocity relation
\begin{equation}
v = \min\!\left(\max\!\left(\frac{s-b_f}{T},0\right),v_0\right).
\end{equation}
Thus, the speed becomes zero at spacing $s=b_f$.

\subsection{Motivation-to-parameter mapping details}

\begin{table}[htbp]
    \centering
    \caption{Motivation-dependent parameter sets and transitions.}
    \label{tab:motivation_parameters}
    \small
    \begin{tabular}{@{}lccc@{}}
        \toprule
        \textbf{Parameter} & \textbf{Low} & \textbf{Normal} & \textbf{High}  \\
        \midrule
        Motivation ($m$) & 0.1 & 1.0 & 3.0 \\
        \\ \addlinespace
        Desired speed ($v_0$) [m/s] & 0.5 & 1.2 & 3.6 \\
        \addlinespace
        Time gap ($T$) [s] & 2.0 & 1.0 & 0.01 \\ \addlinespace
        Buffer ($b$) [m] & 1.0 & 0.1 & 0.0\\ \addlinespace                
        Effective turning bound ($\theta_{\max}$) [rad] & 0.0 & 0.1 & 0.9 \\ \addlinespace
        Neighbor interaction range ($D$) [m] & 0.2 & 0.2 & 0.2 \\ \addlinespace
        \bottomrule
    \end{tabular}
\end{table}

\begin{figure}[htbp]
    \centering
    \includegraphics[width=\textwidth]{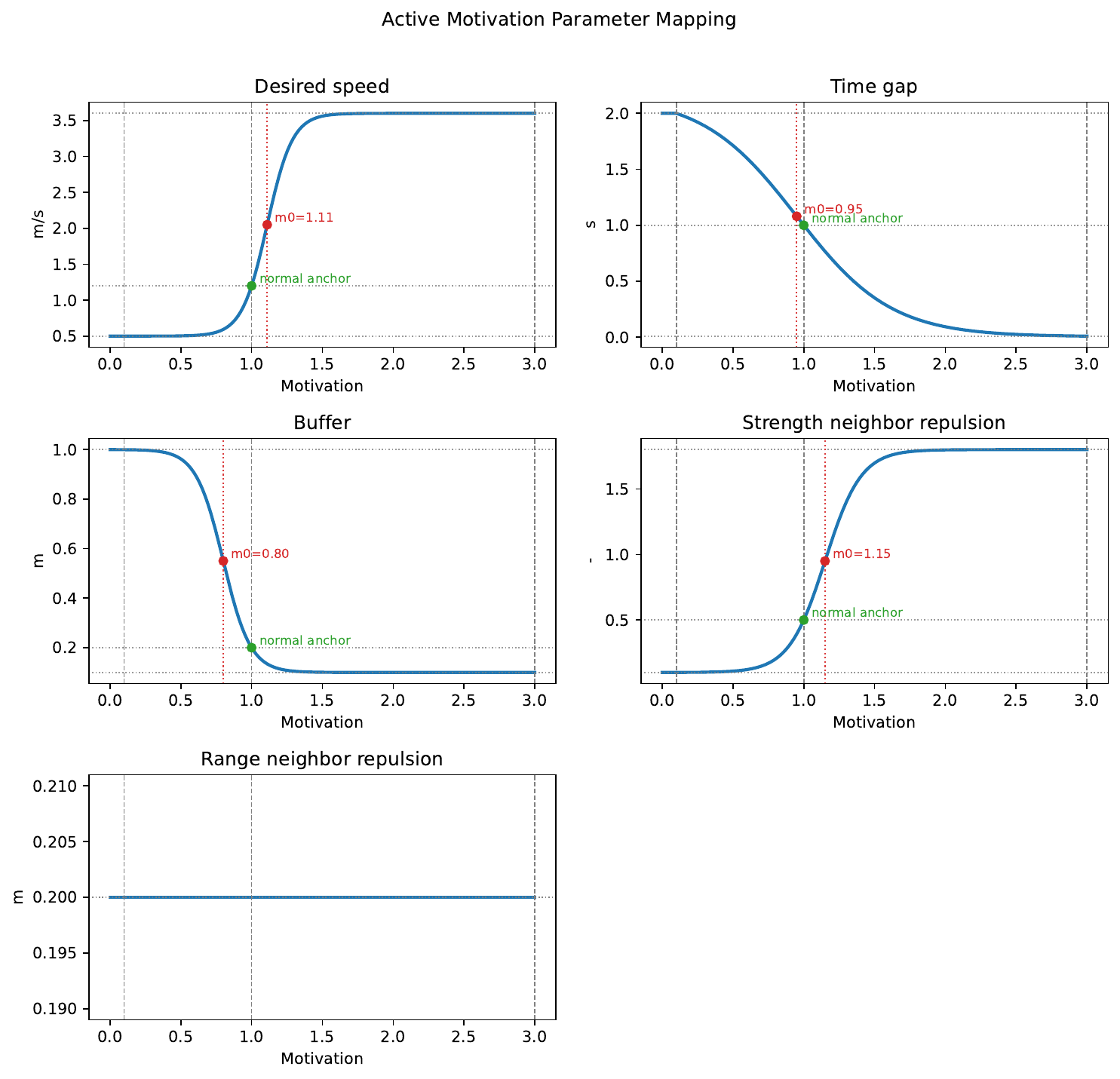}
    \caption{Motivation-to-parameter mappings for all five behavioral parameters, computed from the anchors in \cref{tab:motivation_parameters}. Each panel shows the logistic curve, the inflection point $m_0$ (red), and the normal anchor (green). The range of neighbor repulsion is constant across motivation levels.}
    \label{fig:parameter-mappings}
\end{figure}

\subsection{Submodel ablation: per-agent rank--area scatter}\label{app:submodel-ablation}

\begin{figure}[htbp]
  \centering
  \textbf{$N=80$, closed door}

  \vspace{0.3cm}

  \begin{subfigure}[b]{0.32\textwidth}
    \includegraphics[width=\linewidth]{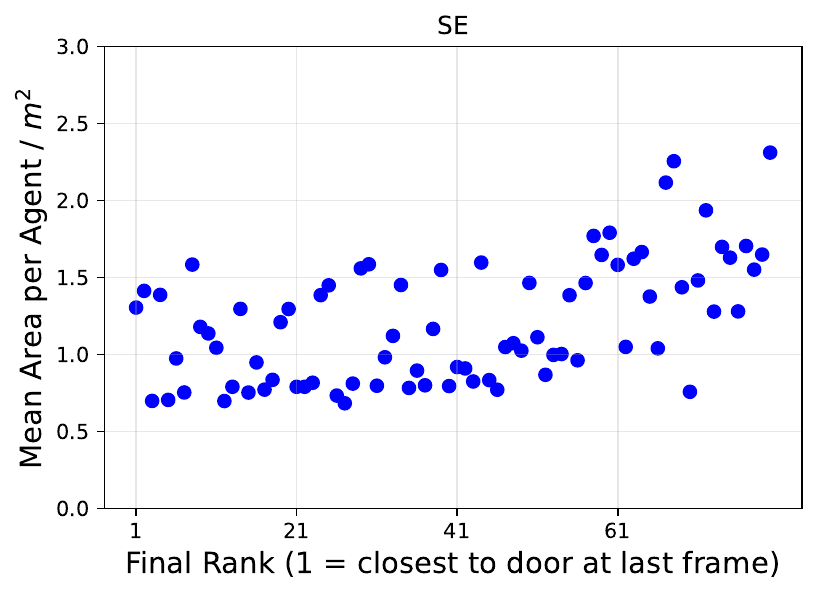}
    \caption{$SE$}
  \end{subfigure}
  \hfill
  \begin{subfigure}[b]{0.32\textwidth}
    \includegraphics[width=\linewidth]{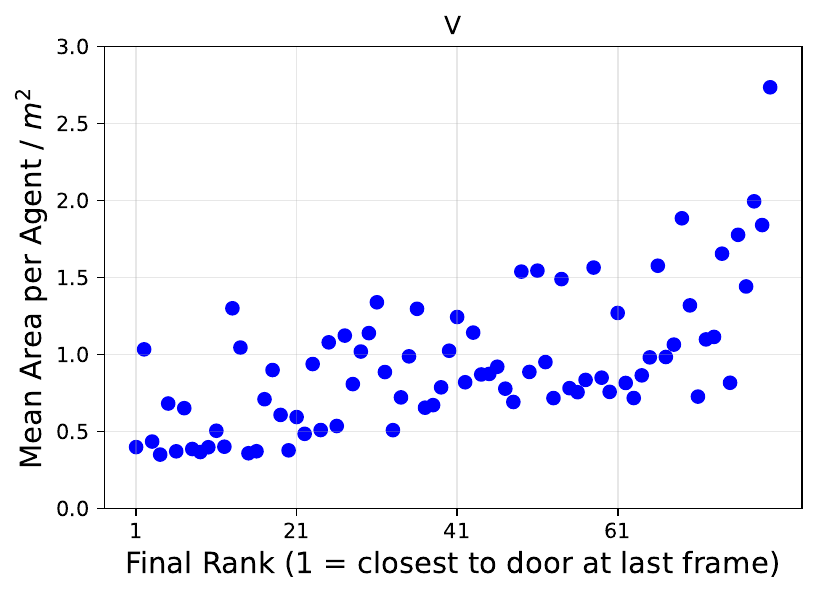}
    \caption{$V$}
  \end{subfigure}
  \hfill
  \begin{subfigure}[b]{0.32\textwidth}
    \includegraphics[width=\linewidth]{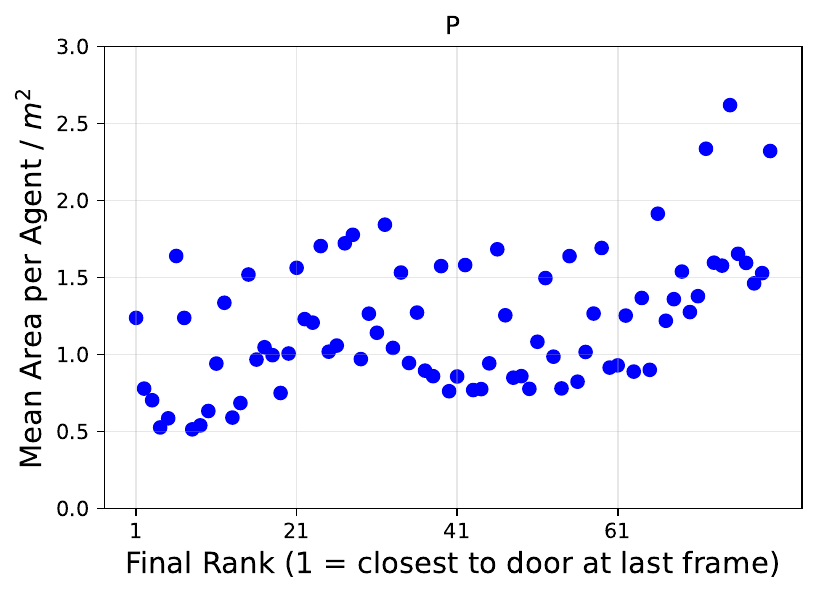}
    \caption{$P$}
  \end{subfigure}

  \vspace{0.4cm}

  \begin{subfigure}[b]{0.45\textwidth}
    \includegraphics[width=\linewidth]{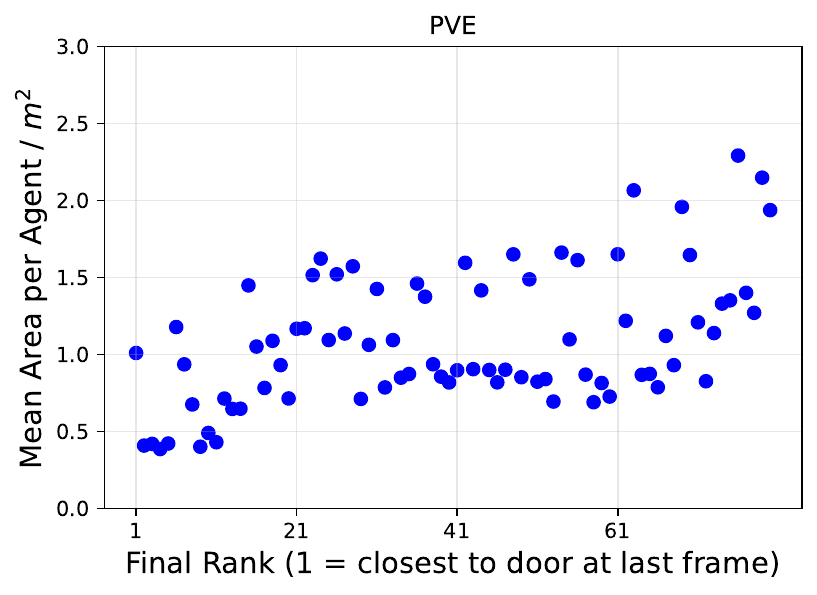}
    \caption{EVP}
  \end{subfigure}
  \hfill
  \begin{subfigure}[b]{0.45\textwidth}
    \includegraphics[width=\linewidth]{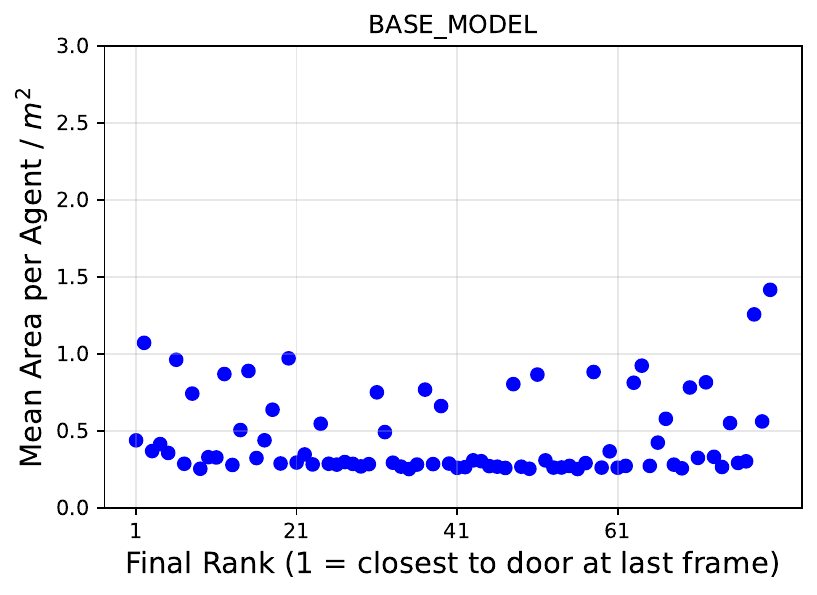}
    \caption{uniform-motivation baseline}
  \end{subfigure}

  \caption{Single-seed per-agent rank--area scatter for each submodel, shown as an ablation of the payoff ($P$), value ($V$), and expectancy ($SE$) terms that compose the combined EVP response. The uniform-motivation baseline is included for reference. The main-text band plots (Figure~\ref{fig:rank-area-bands}) provide the paired-seed statistical comparison of EVP versus base.}
  \label{fig:final-rank-scenarios}
\end{figure}

Figure~\ref{fig:final-rank-scenarios} shows the per-agent rank--area scatter for the individual submodel ablations ($SE$, $V$, $P$) together with the combined EVP response and the uniform-motivation baseline, for both population sizes. Each panel is a single-seed snapshot, shown to expose how the payoff, value, and expectancy contributions shape the rank--area projection individually; the main-text paired-seed band plots (Figure~\ref{fig:rank-area-bands}) summarize the EVP versus base comparison across all ten seeds.

\section{Reflections on the Interdisciplinary Collaboration}\label{app:reflect}
The development of this paper extended over approximately five years and involved numerous iterations of ideas, discussions, and conceptual refinements. Throughout this period, the authors’ value of completing the project remained high, while their expectancy of finishing the project fluctuated over time. A central factor in this dynamic was the gradual development of interdisciplinary communication. In the early stages, even fundamental concepts such as “motivation” were interpreted differently across the authors’ disciplinary backgrounds.

The initial model phase can be characterized by extensive discussions, a non-uniform understanding of motivation, and resulted in an unpublished study within author Üsten's PhD thesis. The model distinguished between high and low motivation and included a spatial segmentation of the environment, in which agents were assigned to high or low motivation levels. These differences were operationalized through variations in desired speed $v_0$ and time to close gaps $T$.

The second phase took place within a broader community effort. During this stage, author Chraibi observed that many of the concepts used in the motivation project were not included in the Glossary for Research on Human Crowd Dynamics \cite{Adrian2019}, an interdisciplinary resource developed to standardize terminology in the multidisciplinary field. This gap motivated Chraibi and Üsten to initiate an extension of the glossary based on their experience. Upon reaching out to colleagues, it became clear that there was broad support for a second edition. Over the course of approximately one year, this effort resulted in a revised and expanded glossary, a second edition \cite{Adrian2025}, involving more than 60 authors worldwide, incorporating around 50 additional concepts and revisions. Within this project, the entry of motivation was developed by author Sieben and her team.

The third phase involved all four authors and introduced expectancy–value theory as the central framework. At this stage, interdisciplinary communication had improved, and the authors made a deliberate effort to understand and use each other’s disciplinary language. The concepts of expectancy and value were jointly interpreted from multiple perspectives: their psychological meaning, their representation in movement parameters (e.g., desired speed), and their relevance in everyday situations. Over time, motivation evolved into a multidimensional construct that integrated definitions from different disciplines.

With improved communication, the focus shifted toward refining the model: making it more realistic, more psychologically grounded, and at the same time more parsimonious. This phase can be characterized by frequent structural revisions. Initially, expectancy–value formulations were extended to include a third component related to competition ($M = EVC$). However, this concept was later reformulated as a payoff structure, reflecting that the competitive context is determined by the underlying structure of rewards rather than by competition per se. Then, expectancy was specified more precisely as including spatial expectancy and payoff structure ($E = SE\cdot P$). Consequent revisions further refined the mathematical structure. The combination of expectancy components was changed from a multiplicative to an additive formulation, as both components represent likelihoods and a multiplicative form led to unintuitive scaling effects ($E = SE+P$). Similarly, the representation of value evolved from a normalized scale to an ordinal scale (Likert-type), better capturing differences in subjective importance in terms of strength ($V = [1, 7]$). Further parameters were introduced ($a$ and $d$) to better capture these dynamics that could not be represented by $v_0$ and $T$ alone.

Upon completing this study, the value of the project remained unchanged for the authors. However, their expectancy of success, which fluctuated over time, has now stabilized at a high level, as the project has reached its final stage. In retrospect, the role of the payoff structure is also reflected in the collaboration itself: progress depended not only on individual effort, but on the evolving structure of shared understanding and collective alignment.

\end{appendices}

\bibliography{sn-bibliography}

\end{document}